\RequirePackage{ifpdf}
\documentclass[12pt,a4paper]{JHEP}
\pdfoutput=1
\usepackage{amssymb,amsfonts}
\usepackage{amsmath}
\usepackage{fancybox}
\usepackage{slashed}
\usepackage{ulem}
\usepackage{color}
\usepackage{cite}
\usepackage{graphicx}
\let\normalcolor\relax

\relax
\def\be{\begin{equation}}
\def\ee{\end{equation}}
\def\bea{\begin{eqnarray}}
\def\eea{\end{eqnarray}}

\newcommand{\bear}{\begin{eqnarray}}

\newcommand{\eear}{\end{eqnarray}}

\def\sq
\def\y{\psi}

\DeclareMathOperator{\Tr}{Tr}

\DeclareMathOperator{\Res}{Res}

\DeclareMathOperator{\cn}{cn}

\DeclareMathOperator{\ns}{ns}
\DeclareMathOperator{\sn}{sn}

\def\bZ {\mathbb{Z}}

\newcommand{\beq}{\begin{equation}}
\newcommand{\eeq}{\end{equation}}
\newcommand{\bal}{\begin{equation}\begin{aligned}}
\newcommand{\eal}{\end{aligned}\end{equation}}
\newcommand{\half}{\frac{1}{2}}

\newcommand{\eqn}[1]{(\ref{#1})}
\newcommand{\nn}{\nonumber}

\newcommand{\cN}{{\mathcal N}}

\newcommand{\cO}{{\mathcal O}}
\newcommand{\cS}{{\mathcal S}}

\newcommand{\cI}{{\mathcal I}}

\def\new#1{{\color [rgb]{0,0,1}#1}}
\title{\boldmath
The $\cN=2$ Schur index from free fermions}

\author{Jun Bourdier, Nadav Drukker and Jan Felix
\\
Department of Mathematics, King's College London,
\\
The Strand, WC2R 2LS, London, United-Kingdom
\\
\email{jun.bourdier@kcl.ac.uk},
\email{nadav.drukker@gmail.com},
\email{jan.felix@kcl.ac.uk}}

\preprint{} 

\abstract{
We study the Schur index of 4-dimensional $\cN=2$ 
circular quiver theories. We show that the index can be expressed as a 
weighted sum over partition functions describing systems 
of free Fermions living on a circle.
For circular $SU(N)$ quivers of arbitrary length we evaluate the large $N$
limit of the index, up to exponentially suppressed corrections. For
the single node theory ($\cN=4$ SYM) and the two node 
quiver we are able to go beyond the large $N$ limit, 
and obtain the complete, all orders large $N$ expansion of the index, 
as well as explicit finite $N$ results in terms of elliptic functions.
}


\begin{document}

\addtolength{\parskip}{.5mm}

\section{Introduction and Results}

Supersymmetric field theories have seen dramatic advances in recent years,
many brought about through the study of partition functions 
on compact manifolds admitting Killing spinors. Most often those are 
the sphere partition function and the superconformal index related to the 
$S^{d-1}\times S^1$ partition function.
The latter, first introduced for 4-dimensional theories in 
\cite{Romelsberger2006,Kinney2007}, is a generalization of the Witten index
\cite{Witten1982}. As such, it counts the states of the theory according to their 
fermionic or bosonic nature, as well as according to their quantum numbers. 
These charges, for symmetries which commute with the Hamiltonian and preserved supercharges, 
are introduced in the usual trace formula through fugacities.

In this paper we study the index of $4d$ $\mathcal{N}=2$ superconformal 
theories on $S^3\times S^1$.
The representations of the corresponding superconformal 
algebra $SU(2,2|2)$ are labelled by the Cartans $(E,j_1,j_2,R,r)$ of 
its bosonic subalgebra. $E$ is the energy, 
$(j_1,j_2)$ are the Cartans of 
the $SU(2)_1 \otimes SU(2)_2$ isometries and $(R,r)$ are 
the Cartans of the 
$SU(2)_R \otimes U(1)_r$ $R$-symmetry group. 

The superconformal index generally has 
three independent fugacities coupling to linear combinations of these Cartans, 
and in addition fugacities for flavour symmetries (see Appendix~\ref{ellipticgamma}).
We are interested in an unrefined version of the index, 
known as 
the Schur index \cite{Gadde2011,Gadde2013,Razamat2012}, 
where one relation is imposed between the three fugacities, 
but it turns out that the resulting index depends only on one unique fugacity $q$. 
The charge that this fugacity couples to commutes with a pair of supercharges $Q$ and $Q'$, 
which, following \cite{Gadde2013, Gadde:2011ia}, we choose 
such that
\bal
\delta &\equiv 2 \{Q, Q^\dagger \} = E -2j_2 - 2R + r \,, \\
\delta ' &\equiv 2 \{Q', {Q'}^\dagger \} = E +2j_1 - 2R - r\,.
\eal
The Schur index is then given by%
\footnote{We use a slightly different definition of the fugacity than in most of the literature. 
$q$ in \cite{Razamat2012} corresponds to $q^2$ in our notations.}
\beq
\label{4dindex}
\mathcal{I} = \Tr(-1)^F 
e^{\beta \delta}
e^{\beta' \delta'}
q^{2 \left(E - R\right)} 
\prod_a e^{2iu^{(a)}F^{(a)}}
\,,
\eeq
where $q$
and $e^{2iu^{(a)}}$ are the fugacities for the charges of the superconformal 
and flavour groups respectively, and $F^{(a)}$ are the flavour charges.
We express the flavour fugacities in terms of their chemical potentials
$u^{(a)}$,
which 
appear in a natural way 
in the explicit expressions for the index below. 
The sum 
in \eqn{4dindex} 
is taken over all states of the theory, but following the usual Witten 
argument \cite{Witten1982}, contributions from fermions and bosons cancel 
for all multiplets except for those with 
$\delta = \delta' =0$, so that the index is independent of $\beta,\beta'$.

As shown in \cite{Romelsberger2006, Kinney2007}, 
there is an elegant way to express the 
index \eqref{4dindex} as a matrix model. 
It was first noted in \cite{Dolan2009} that the contributions 
from each multiplet
can be neatly expressed in terms of elliptic gamma functions. 
For the Schur index, the contributions combine in such a way that they can be written as 
$q$-theta functions (see Appendix~\ref{ellipticgamma} and \cite{Razamat2012}).
We give here the expressions for each multiplet, 
in terms of Jacobi theta functions and the Dedekind eta function 
(see Appendix~\ref{identities} 
for their definition and for useful identities). 
We consider $SU(N)$ gauge 
groups, and use the gauge freedom to reduce the integral 
over the Lie algebra to an integral over a Cartan subalgebra that 
we parametrize with $\alpha_i$, $i=1, \cdots, N$, which are $\pi$ 
periodic and which satisfy the traceless condition $\sum_{i=1}^N \alpha_i =0$ 
imposed through a delta function.

The contribution from an $\mathcal{N}=2$ vector multiplet, 
including the integration with the 
Haar measure is \eqn{N2vec}
\beq
\label{4dvec}
\mathcal{I}_{\text{vec}}
=
\big(q^{-\frac{1}{12}} \eta(\tau)\big)^{2(N-1)}
\frac{1}{\pi^N N!}\int_{0}^{\pi} d^N \alpha 
\delta\left( \sum_{i=1}^N \alpha_i \right)
\prod_{i <j} 
\frac{\vartheta_1^2\big(\alpha_i-\alpha_j, q \big)}
{q^{\frac{1}{3}}\eta^{2}(\tau)} \, ,
\eeq
where we have used the notation $q \equiv e^{i\pi\tau}$.

An $\mathcal{N}=2$ hypermultiplet in the bi-fundamental representation 
of the gauge groups $G^{(1)} \times G^{(2)}$ contributes to the index as \eqn{n2hyp}
\beq
\label{4dhypbif}
\mathcal{I}_{\text{hyp}}
= \prod_{i,j} 
\frac{q^{-1/12}\eta(\tau)}
{\vartheta_4\big(\alpha^{(1)}_i-\alpha^{(2)}_j + u, q \big)}\,,
\eeq
where $u$ is the chemical potential for the $U(1)$ flavour symmetry,
which enters in the index as defined in \eqn{4dindex}. 

The simplest matrix model of this class corresponds to $\cN =4$ SYM, 
which in this $\cN =2 $ formalism has one $U(N)$ gauge group, 
as well as one adjoint hypermultiplet. 
The model for this particular case was solved exactly 
in \cite{Bourdier:2015wda}. 
This was made possible by expressing the matrix model as 
the partition function of a 1-dimensional free Fermi gas. 
In the context of supersymmetric field theories, this type of manipulation 
was pioneered in \cite{Marino2012}, who studied the $S^3$ partition function of ABJM 
theory 
\cite{Aharony2008}, as well as more 
general circular quiver gauge theories. This led to numerous results, 
such as the discovery of the universal 
Airy function behaviour of the perturbative part in the large $N$ expansion 
for all circular quivers \cite{Marino2012}, as well as a complete understanding of the 
non-perturbative effects of the ABJ(M) partition function 
(see \cite{Hatsuda:2015gca} for a review). 
This formalism was also successfully 
applied to many other three dimensional superconformal theories, 
with wide ranging gauge groups \cite{Mezei2014} 
and quiver structures \cite{Assel:2015hsa,Moriyama2015}, 
and was also used to understand the relationship between 
topological strings and 3d partition functions
\cite{Hatsuda2014a,Huang:2014eha,Grassi2014a}.

The key step in this method is 
the 
use 
of 
a determinant identity which expresses 
the integrand of the matrix model as a determinant. 
Indeed, in \cite{Bourdier:2015wda} we used 
an elliptic generalization \eqref{jacobidet} of the Cauchy determinant 
identity which resolved the interactions in the matrix model of the index of 
$\cN=4$ SYM.

\begin{figure}[ht]
\label{fig:quiver}
\centering
\includegraphics[width=0.5\textwidth]{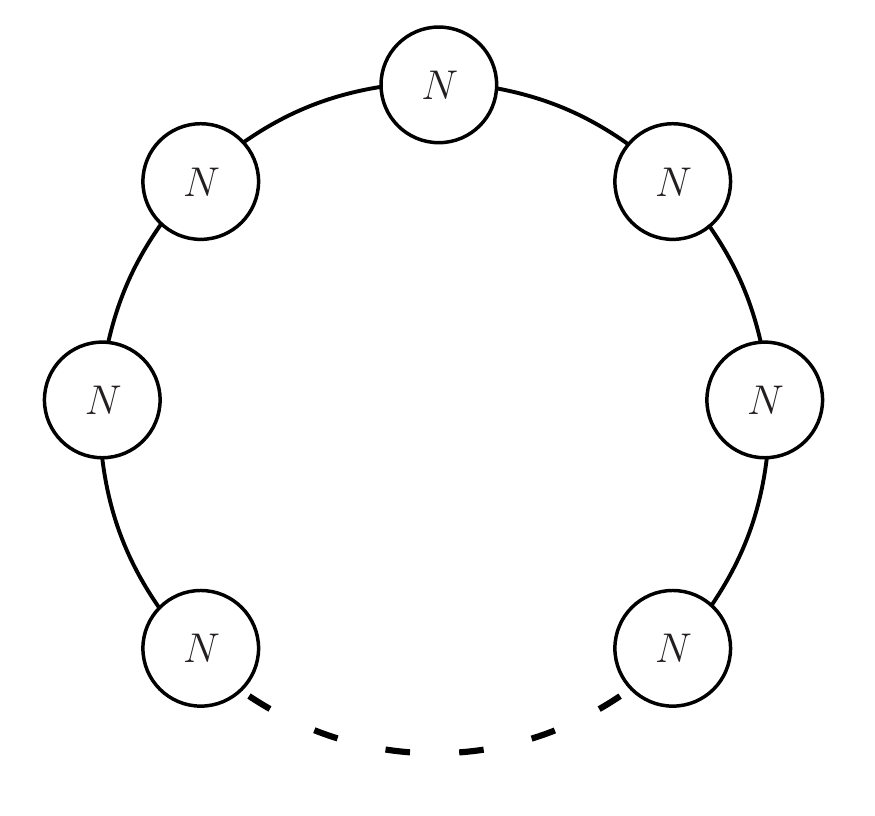}
\caption{A diagram of $\cN=2$ circular quiver theories. 
Each node of the diagram represents a $U(N)$ 
$SU(N)$ vector multiplet, while a solid lines connecting two nodes represents
a bi-fundamental hypermultiplet.}
\end{figure}

In the present paper we study theories that are a natural generalisation 
of $\cN=4$ SYM, namely circular quivers, and apply 
the Fermi gas formalism to compute the Schur index. 
A circular quiver of length $L$ has gauge group $SU(N)^L$,
with vector multiplets for each gauge group factor
and bi-fundamental hypermultiplets connecting them in circular fashion,
as depicted in figure~\ref{fig:quiver}. These theories are of particular 
interest as the circular structure is most susceptible to a Fermi-gas 
interpretation in terms of traces of single particle density operators 
(theories with a quiver structure of a $\hat D$ Dynkin diagram 
are also interesting candidates,
and a discussion of these will appear in \cite{JunD}). 
It would be interesting to understand whether any of the techniques 
developed here could also be applied to non-Lagrangian ``class-$\cS$'' theories.

Unlike 3d circular quivers, where many theories with $\cN\geq2$ flow to conformal 
fixed points, in 4d we cannot add extra fundamental hypermultiplets. The resulting 
theories are neither conformal nor asymptotically free. Nonetheless, it is rather 
easy to add to the matrix models factors associated to fundamental matter, 
but since we
do not have a 4d interpretation of this, we do not consider these here.%
\footnote{Supersymmetric partition functions and indices have been calculated 
for non-renormalizable theories, including gauge theories in $d>4$ and supergravity 
theories. So there may yet be a meaning for the index of 4d theories with positive 
$\beta$-functions.}
Note however that in the calculation of the Schur index of $\cN=4$ SYM in the presence of 
Wilson loop operators \cite{Drukker:2015spa}, the matrix model gets enriched by terms 
somewhat similar to those due to fundamental matter fields. It would be interesting to try 
to generalize that calculation to the circular quivers studied here and explore this 
generalization of the matrix model.

We find that for $SU(N)$ circular quivers 
one can use the elliptic determinant identity 
to write the index as the partition function describing a set of 
$N$ \emph{interacting} fermions. The interaction terms are due 
to the tracelessness condition,which we avoid handling directly 
by expanding them in Fourier series, 
at the cost of introducing an \emph{infinite} number of terms, 
each of which can be interpreted as the partition function of a 
\emph{free} Fermi gas. 

The resulting infinite sum has a natural interpretation as a Fourier 
expansion in flavour 
fugacities of a rescaled index 
(provided the product of the flavour fugacities is $1$). Thus each 
Fourier coefficient of the rescaled index is given 
by the aforementioned partition function of a free Fermi gas \eqref{ISUN}. 
Each Fermi gas can then be studied independently, 
and we do so by considering the associated grand canonical partition function 
with chemical potential $\mu$. 

For a circular quiver of arbitrary length we are able to write down 
(when the product of the flavour fugacities is $1$) 
a closed formula for the grand partition function \eqref{Xi-L-theta}, 
involving a product of Jacobi elliptic theta functions 
evaluated at the roots of a polynomial \eqref{polynomialdi}, whose degree 
grows with the length of the quiver. 

In Section~\ref{sec:LargeN} 
we then present a computation which gives the 
full perturbative (in $N$) expression for 
the index in the large $N$ limit. We first calculate the leading term 
of the grand canonical partition function at large chemical potential $\mu$, 
for which we give two different methods. 
In one method we solve for the roots of 
the polynomial \eqref{polynomialdi} in the large $\mu$ limit, 
while in the other we use a Mellin-Barnes 
representation of the grand potential to extract its large $\mu$ behaviour.
We can then carry out the resummation over the 
Fourier modes to obtain
\beq
\label{largeNresult}
\mathcal{I}(N,L)
= \frac{q^{\frac{L}{6}}} 
{\eta^{L}(\tau)\eta^2(\tfrac{L\tau}{2})} + \cO(e^{-cN} )\,, 
\qquad 
c>0\,.
\eeq
We find here that the leading term is $N$-independent, as was already pointed out in 
\cite{Kinney2007} for $\mathcal{N}=4$ SYM, 
that there are no perturbative $1/N$ corrections, 
and that 
the result 
is also independent of the flavour fugacities.

It should be noted that this scaling does not match the supergravity action, 
which grows as $N^2$. Indeed the correct quantity to compare on the 
field theory side to the classical supergravity action is the partition 
function on $S^3\times S^1$, which is related to our index through a factor, 
dubbed the supersymmetric Casimir energy \cite{Assel2014,Assel2015}, 
which does scale like $N^2$. There is still a discrepancy with the supergravity calculation, 
possibly due to missing counterterms of supersymmetric holographic renormalization. 
As we see above, we find the same $N^0$ scaling for the index of $\cN =2$ theories, 
and have not calculated the Casimir energy. It is unclear whether the $\cN=2$ 
generalization could shed light on this issue.

In the cases where we can find the non-perturbative corrections at large $N$, which 
are 
$\cN=4$ SYM \cite{Bourdier:2015wda} and the two node quiver 
(see Section~\ref{sec:exactlargeN}), there should be an interpretation of the instanton 
corrections in terms of some other supergravity saddle points and/or D-brane configurations, 
but such an understanding is also lacking.

As just mentioned, for short quivers we are able to go beyond the
perturbative 
large $N$ result
\eqn{largeNresult}. This is based on the expression for 
the grand partition function in terms of the roots of a polynomial, which for 
$L=1$ and $L=2$ is quadratic, so the roots can be obtained explicitly. 
This allows us to compute the complete, 
all orders large $N$ expansions of the index for these theories in Section~\ref{sec:exactlargeN}.
For the two-node case, 
we find that the 
independence of the asymptotic expression \eqn{largeNresult} 
on the flavour fugacity
gets lifted by non-perturbative corrections.
Furthermore, 
in the absence of 
flavour fugacities, 
we extract exact results in terms
of elliptic integrals for finite values of $N$ in Section~\ref{sec:finite}. 
Finally in Section~\ref{sec:q} we are able to 
obtain similar finite $N$ results also for longer quivers by comparing the 
$q$ expansion of the index and polynomials of elliptic integrals, and 
present results for quivers of up to four nodes.

\section{Circular quivers as free Fermi gases}
\label{sec:fermi}

Using the formulae given in the introduction, the matrix model for the Schur 
index of a $SU(N)$ circular quiver gauge theory with $L$ nodes is
\bal
\label{indexqtheta}
\mathcal{I}(N,L)
&=\frac{q^{-\frac{LN^2}{4}} q^{\frac{L}{6}}\eta(\tau)^{(3N-2)L}}{N!^L \pi^{NL}}
\int_{0}^{\pi} \prod_{a=1}^L d^N\alpha^{(a)} 
\delta\left( N A^{(L)} \right) 
\prod_{a=1}^{L-1} \delta \big( N \big( A^{(a)} - A^{(a+1)} \big) \big)
\\
& \hskip2in \times
\frac{\prod_{i<j} 
\vartheta_1\big(\alpha^{(a)}_i-\alpha^{(a)}_j \big)
\vartheta_1\big(\alpha^{(a+1)}_i-\alpha^{(a+1)}_j \big)}
{\prod_{i,j} \vartheta_4\big(\alpha^{(a)}_i-\alpha^{(a+1)}_j + u^{(a)} \big)}\,,
\eal
where it is understood that $\alpha^{(L+1)}_i = \alpha^{(1)}_i$, and we have used the notation 
$\vartheta_i(z)= \vartheta_i(z,q)$, and introduced the centers of mass
\beq
\label{za}
A^{(a)}=\frac1N \sum_{i=1}^N \alpha_i^{(a)} \,.
\eeq  
We allow here for arbitrary chemical potentials 
$u^{(a)}$ for the flavour symmetries of the hypermultiplets.
The $L$ delta function constraints in 
\eqref{indexqtheta} come from the tracelessness condition, and we have chosen 
to represent all but one of them in difference form. 
Note that for the single node theory ($\mathcal{N} =4$ SYM) the traceless 
condition also applies to the hypermultiplet, where the adjoint is of dimension
$N^2-1$ rather than $N^2$. This introduces and additional factor of 
$q^{\frac{1}{12}}\eta^{-1}(\tau)\vartheta_4(u)$.

The second line in \eqn{indexqtheta} can be rewritten using 
an elliptic determinant identity \eqn{jacobidet} given in 
Appendix~\ref{sec:elipticcauchy}
\bal
& \frac{\prod_{i<j} 
\vartheta_1\big(\alpha^{(a)}_i-\alpha^{(a)}_j \big)
\vartheta_1\big(\alpha^{(a+1)}_i-\alpha^{(a+1)}_j \big)}
{\prod_{i,j} \vartheta_4\big(\alpha^{(a)}_i-\alpha^{(a+1)}_j + u^{(a)} \big)}
=\left(\frac{\vartheta_2}{\vartheta_4\vartheta_3}\right)^N 
\\& \hspace{.5cm}\times
\frac{\vartheta_3\,q^{-\frac{N^2}{4}} e^{iN^2(A^{(a)}-A^{(a+1)}+u^{(a)})}}
{\vartheta_3\big(N(A^{(a)}-A^{(a+1)}+u^{(a)} + \tfrac{\pi\tau}{2})\big)} 
\det_{ij} \left(\cn \big(\big(\alpha_i^{(a)} -
\alpha_j^{(a+1)} +u^{(a)} \big)\vartheta_3^2 \big)\right) 
\, ,
\eal
where we have used the notation $\vartheta_i = \vartheta_i(0,q)$.

Putting this together allows us to write the index as
\beq
\label{jacobiindexsun}
\mathcal{I} (N,L)
=\frac{q^{- \frac{N^2 L}{2}} e^{- i N^2 U} q^{\frac{L}{6}}}
{\eta^{2L}(\tau) }
\prod_{a=1}^L
\frac{\vartheta_3}{\vartheta_3 \big(N u^{(a)}+ N\frac{\pi\tau}{2} \big)}
\,Z(N)
\eeq
where $U=\sum_{a=1}^Lu^{(a)}$ and where we have used the identity 
\eqref{reduction}
to simplify the $\alpha$-independent factor. $Z(N)$ is defined as%
\footnote{The factor of $\vartheta_2^2/2\pi$ is 
included with the cn functions to simplify their Fourier expansion below.}
\bal
Z(N, L)&= \frac{1}{N!^L}\int_{0}^{\pi} 
\prod_{a=1}^Ld^N\alpha^{(a)}
\delta \left( NA^{(L)} \right) \prod_{a=1}^{L-1} \delta \left( N \left( A^{(a)} - A^{(a+1)} \right) \right)
\\
&\quad{}\times
\det_{ij} \bigg(  
\frac{\vartheta_2^2}{2 \pi} \cn \big((\alpha_i^{(a)} 
- \alpha_j^{(a+1)} +u^{(a)})\vartheta_3^2 \big)\bigg),
\eal

Each determinant can then be written as a sum over permutations, and by relabelling the eigenvalues, 
one can factor all but one of the permutations, picking up a factor of $N!^{L-1}$ and leading to
\begin{align}
Z&(N, L )= \frac{1}{N!} 
\sum_{\sigma \in S^N} 
(-1)^{\epsilon(\sigma)}\int_{0}^{\pi} \prod_{a=1}^L d^N\alpha^{(a)} \,
\delta\left(N A^{(L)} \right) \prod_{a=1}^{L-1} \delta \big( N \big(A^{(a)} - A^{(a+1)} \big) \big)
\\&\quad{}
\label{jacobiindex}
\times 
\prod_{i=1}^N \bigg(\prod_{a=1}^{L-1} 
\frac{\vartheta_2^2}{2 \pi} 
\cn \big((\alpha_i^{(a)} - \alpha_i^{(a+1)} +u^{(a)})\vartheta_3^2 \big)\bigg)
 \frac{\vartheta_2^2}{2 \pi} 
\cn \big((\alpha_i^{(L)} - \alpha_{\sigma(i)}^{(1)} +u^{(L)})\vartheta_3^2 \big),
\nn
\end{align}
This 
expression strongly suggests that the eigenvalues 
describe fermionic degrees of freedom. The difficulty in writing down 
a single particle density operator comes 
from the presence of the center of mass coordinates $A^{(a)}$ 
in the delta functions, 
which introduce complicated interactions.

To overcome it we first shift the eigenvalues as 
$\alpha^{(a)} \rightarrow \alpha^{(a)} + \sum_{b=1}^{a-1} u^{(b)}$ 
so that the $u$'s appear only inside the delta functions and one $\cn$
\bal
\label{ZN}
Z(N,L)&=
\sum_{\sigma \in S^N} 
\left(-1 \right)^{\epsilon(\sigma)}
\int_{0}^{\pi}
\prod_{a=1}^{L-1}d^N\alpha^{(a)} \delta\big(N(A^{(a)} - A^{(a+1)} - u^{(a)} )\big)
\\&\quad{}\times 
\int_{0}^{\pi}d^N\alpha^{(L)} 
\delta\big(N (A^{(L)} + U - u^{(L)} ) \big)
\\&\quad{}\times 
\prod_{i=1}^N \bigg(
\frac{\vartheta_2^2}{2 \pi} 
\cn \big((\alpha_i^{(L)} - \alpha_{\sigma(i)}^{(1)} + U)\vartheta_3^2 \big)
\prod_{a=1}^{L-1} 
\frac{\vartheta_2^2}{2 \pi} 
\cn \big((\alpha_i^{(a)} - \alpha_i^{(a+1)} )\vartheta_3^2 \big)
\bigg),
\eal
Now we represent the delta functions by their Fourier expansion
\bal
\delta
&
\big(N (A^{(L)} + U -u^{(L)}) \big)
\prod_{a=1}^{L-1}
\delta\big(N(A^{(a)}-A^{(a+1)}- u^{(a)})\big)
\\
&
=
\sum_{\vec{n} \in \bZ^L}
e^{- 2i N \sum_{a=1}^{L} n^{(a)} u^{(a)}}
e^{2 i N U n^{(L)}}
e^{2 i n^{(L)}\sum_i\alpha^{(L)}_i} 
\prod_{a=1}^{L-1} 
e^{2 i n^{(a)}\sum_i(\alpha^{(a)}_i -\alpha^{(a+1)}_i)}
\,,
\eal
with $\vec{n} = \{n^{(a)} \}$. 
We can then write the rescaled index as a sum
\beq
\label{ISUN}
Z(N, L )
=\sum_{\vec{n} \in \bZ^L} 
e^{- 2i N \sum_{a=1}^{L} n^{(a)} u^{(a)}}
e^{2 i N U n^{(L)}}
Z_{\vec{n}} \,.
\eeq
Now each Fourier coefficient
$Z_{\vec{n}}$ is a partition function of a {\em free} Fermi gas expressed as
\begin{align}
\label{Inasun}
Z_{\vec{n}} &= 
\frac{1}{N!}\sum_{\sigma \in S^N} (-1)^{\epsilon(\sigma)}
\int_{0}^{\pi} 
d^N\alpha_i\prod_{i}^N \rho_{\vec{n}}(\alpha_i,\alpha_{\sigma(i)} )\,,
\end{align}
in terms of a single particle density operator
\bal
\label{rhonasun}
&\rho_{\vec{n}}\big(\alpha^{(1)},\alpha^{(L+1)}\big)
=\int_{0}^{\pi} \prod_{a=2}^Ld\alpha^{(a)} \,
e^{2 i n^{(L)} \alpha^{(L)}} 
\frac{\vartheta_2^2}{2 \pi}
\prod_{a=1}^{L-1} 
e^{2 i n^{(a)}(\alpha^{(a)} - \alpha^{(a+1)})}
\\
& \hskip2cm \times 
\cn \big((\alpha^{(L)} - \alpha^{(L+1)} + U)\vartheta_3^2 \big)
\prod_{a=1}^{L-1} 
\frac{\vartheta_2^2}{2 \pi}
\cn \big((\alpha^{(a)} - \alpha^{(a+1)} )\vartheta_3^2 \big)
\,, 
\eal
where in $Z_{\vec{n}}$ we substitute $\alpha^{(1)} = \alpha_i$ 
and $\alpha^{(L+1)} \equiv \alpha_{\sigma(i)} $. 
In fact, this Fourier expansion 
closely
mirrors the original definition of the index with flavour fugacities \eqn{4dindex} 
and were it not for the $u$ dependence in the rescaling factor in \eqn{jacobiindexsun}, 
then $Z_{\vec n}$  would be 
the index for fixed flavour charges $F^{(a)}=-Nn^{(a)}$.

The Fermi gas partition function 
\eqn{Inasun} 
is completely determined by
\bal
\label{Zldef}
Z_{\vec{n};\ell} = \Tr(\rho_{\vec{n}}^\ell)= \int_{0}^{\pi} d x_1 \cdots dx_\ell \,
\rho_{\vec{n}}(x_1, x_2 )\cdots \rho_{\vec{n}} (x_\ell, x_1 )\,,
\eal
often referred to as the \emph{spectral traces}.
Indeed, conjugacy classes of $S_N$ have $m_\ell$ cycles of length $\ell$, 
and from the definition \eqref{Inasun} of $Z_{\vec{n}}(N)$ 
we get
\bal
\label{ZNZl}
Z_{\vec{n}}(N)= {\sum_{\{m_\ell\}}}^\prime 
\prod_\ell 
\frac{Z_{\vec{n};\ell}^{m_\ell} \,(-1)^{(\ell-1)m_\ell}} {m_\ell !\,\ell^{m_\ell}}\,,
\eal
where the prime 
denotes a sum over sets that satisfy $\sum_\ell \ell m_\ell = N $. 

To evaluate $Z_{\vec{n};\ell}$, we first simplify the expression 
for the density $\rho_{\vec{n}}$ \eqref{rhonasun} 
by using the Fourier expansion of the elliptic function
\beq
\cn \big(z\,\vartheta_3^2\big)= \frac{1}{\vartheta_2^2} 
\sum_{p \in \bZ} \frac{e^{i(2p-1)z}}{\cosh i\pi\tau(p - \half)}\,,
\eeq
and we obtain
\begin{align}
\rho_{\vec{n}} &= \sum_{\vec{p}\in\bZ^L}
e^{i(2p^{(L)}-1)U}
\prod_{a=1}^L \frac{1}{2 \pi}
\frac{1}{\cosh i\pi\tau(p^{(a)}-\half)}
\\&\quad{}\times 
\int_{0}^{\pi} \prod_{a=2}^Ld\alpha^{(a)} \,
e^{2 i n^{(L)}\alpha^{(L)}}
e^{2 i(p^{(L)} - \half)(\alpha^{(L)} - \alpha^{(L+1)})}
\prod_{a=1}^{L-1} 
e^{2 i(n^{(a)} + p^{(a)} - \half)(\alpha^{(a)} - \alpha^{(a+1)})}.
\nn
\end{align}
Shifting the summation over $p^{(a)} \rightarrow p^{(a)} - n^{(a)}$, 
and doing the integration over the 
$\alpha^{(a)}$'s
gives
\beq
\rho_{\vec{n}}=
\frac{1}{\pi} \sum_{p \in\bZ}
e^{2i(p -n^{(L)}-\half)U}
e^{2 i(p - \half)\alpha^{(1)}} e^{- 2 i(p - n^{(L)} - \half)
\alpha^{(L+1)}} 
\prod_{a=1}^L\frac{1}
{2 \cosh i\pi\tau \big(p - n^{(a)}-\half\big)} 
\,.
\eeq

As explained above, we are interested in computing the quantity 
$Z_{\vec{n};\ell}$ \eqref{Zldef}. For $\ell=1$ we find
\beq
Z_{\vec{n};1} 
=\int_{0}^{\pi} d\alpha \,
\rho_{\vec{n}}(\alpha,\alpha)
=\delta_{n^{(L)}}
\sum_{p\in\bZ}e^{2i(p -n^{(L)}-\half)U}
\prod_{a=1}^L\frac{1}{2\cosh i\pi\tau \big(p-n^{(a)} - \half\big)}\,.
\eeq
This structure persists also when considering the convolution of several $\rho$'s, with 
a constraint on $n^{(L)}$ and a single sum over $p$
\beq
\label{SUNZl}
Z_{\vec{n};\ell}
=\delta_{n^{(L)}}
\sum_{p\in\bZ}
e^{2i(p -n^{(L)}-\half)U \ell}
\prod_{a=1}^L 
\bigg(\frac{1}{2 \cosh i\pi\tau \big(p-n^{(a)} - \half\big)}\bigg)^\ell
\,.
\eeq
The presence of the $\delta_{n^{(L)}}$ factor in the expression above 
tells us that the sum in \eqref{ISUN} is in reality 
only over $\{n^{(a)} \} \in \bZ^{L-1}$ with $n^{(L)}=0$. 
From now on we omit this Kronecker delta, and the modes $n^{(a)}$ 
run over $a=1, \cdots,L-1$.

We can plug the expressions \eqref{SUNZl} into \eqn{ZNZl} 
to evaluate $Z_{\vec{n}}(N)$ 
and then sum over the integers $\vec{n} \in \bZ^{L-1}$
to find the index $\cI(N,L)$ 
\eqn{jacobiindexsun}, \eqn{ISUN}.
An alternative, which avoids the combinatorics in \eqn{ZNZl} 
is to sum over the indices 
of 
quivers  
with 
arbitrary ranks $N$. 
For each $\vec{n} \in \bZ^{L-1}$, 
we define the associated grand canonical partition function
\beq
\label{grandpartition}
\Xi_{\vec{n}}(\kappa)=1+ \sum_{N=1}^\infty Z_{\vec{n}}(N)\kappa^N \,.
\eeq
$\kappa$ is the fugacity and we write it also in terms of the chemical potential 
$\mu$ as $\kappa=e^\mu$. This definition 
is easily inverted to recover $Z_{\vec{n}}(N)$
\beq
\label{invertXin}
Z_{\vec{n}}(N) = \frac{1}{2 \pi i} 
\int_{-i\pi}^{i\pi} d \mu \,
\Xi_{\vec{n}}(e^\mu)e^{- \mu N} \,.
\eeq

The combinatorics simplify when considering the 
grand potential
\beq
\label{Jmu}
J_{\vec{n}}(\mu)\equiv 
\log\Xi_{\vec{n}}\big(e^\mu \big)
= -\sum_{\ell= 1}^\infty \frac{(-1)^\ell Z_{\vec{n};\ell} e^{\mu \ell}}{\ell}\,,
\eeq
and we
can then easily sum over $\ell$ and find a very compact expression 
\beq
\label{SUNXi}
\Xi_{\vec{n}}(\kappa)
=\prod_{p \in \bZ} \bigg(1 + \kappa e^{2i(p -n^{(L)}-\half)U}
\prod_{a=1}^L 
\frac{1}{2 \cosh i\pi\tau \big(p- n^{(a)}- \half\big)} \bigg)
\,.
\eeq

From now on we focus on the case with $U=0$, {\it i.e.} the product of the flavour fugacities 
is $1$. This allows us to write $\Xi_{\vec{n}}$ 
as a product of theta functions evaluated at the roots of a polynomial.
Indeed, for $U=0$,
each term in the product over $p$ in \eqref{SUNXi} can be written as
\beq
\label{one-term}
\frac{X^L q^{- n}\,\kappa
+ \prod_{a=1}^L\big(1+X^2q^{-2n^{(a)}}\big)}
{\prod_{a=1}^L\big(1+X^2q^{-2n^{(a)}}\big)} \,,
\qquad
X\equiv q^{p-\half}\,,
\eeq
where $n = \sum_{a=1}^{L} n^{(a)}$. 
The numerator of \eqref{one-term} is a polynomial of 
degree $2L$ in $X$ with coefficients that depend on 
$q$, $n^{(a)}$ and $\kappa$, but not on $p$. 
It can be factored as
\beq
\label{defdi}
\frac{\prod_{j=1}^{2L}\big(1+e^{2id_j}X\big)}
{\prod_{a=1}^L\big(1+ X^2q^{-2n^{(a)}}\big)} \,.
\eeq
Now take the term in \eqn{SUNXi} with $p\to-p+1$. We can write it also 
as \eqn{one-term} with the same denominator. The numerator is then of a similar 
form with $n^{(a)}\to-n^{(a)}$, which is factorized by the inverse roots $-e^{2id_j}$.
Splitting then the product in \eqn{SUNXi} over only positive $p$ 
gives 
\bal
\label{Xi-L-theta}
\Xi_{\vec{n}}
&=\prod_{p =1}^\infty
\frac{\prod_{j=1}^{2L} \big(1+e^{2id_j}q^{p-\half}\big)\big(1+e^{-2id_j}q^{p-\half}\big)}
{\prod_{a=1}^L \big(1+q^{-2n^{(a)}}q^{2p-1}\big)\big(1+q^{2n^{(a)}}q^{2p-1}\big)}
=
\frac{\prod_{j=1}^{2L} \vartheta_3\big(d_j,q^{\frac12}\big)}
{\vartheta_4^L\prod_{a=1}^L \vartheta_3(n^{(a)} \pi\tau,q)}
\\&=
\frac{q^{\sum_{a=1}^L(n^{(a)})^2}}{\vartheta_4^L\vartheta_3^L}
\prod_{j=1}^{2L} \vartheta_3\big(d_j,q^{\frac12}\big).
\eal

We cannot find the explicit roots of a polynomial of arbitrary degree, but for $L=1$, it is 
quadratic, which is what allowed us to solve the index for $\cN=4$ SYM in closed form 
\cite{Bourdier:2015wda} (see Section~\ref{sec:N4} below). In fact, for even $L$ 
the numerator of \eqref{one-term} can be viewed as a polynomial of degree 
$L$ in $X^2$, which we use in Section~\ref{sec:L2} to solve the index of the 
two node quiver.

Explicitly, equation \eqn{defdi} factorized into $L$ terms is
\beq
\label{defditilde}
\frac{\prod_{j=1}^{L}\big(1+e^{2i\tilde d_j}X^2\big)}
{\prod_{a=1}^L\big(1+X^2q^{-2n^{(a)}}\big)} \,,
\eeq
and the grand partition function is now expressed in terms of theta functions 
with nome $q$ rather than $q^{\frac{1}{2}}$
\beq
\label{Xi-evenL-theta}
\Xi^\text{even $L$}_{\vec{n}}
=\frac{q^{\sum_{a=1}^L(n^{(a)})^2}}{\vartheta_3^L}
\prod_{j=1}^{L} \vartheta_3(\tilde d_j)\,.
\eeq
Clearly the $2L$ roots for $X$ are given in terms of the new ones by the 
pairs $\pm e^{-i\tilde d_j}$ and the expressions \eqn{Xi-L-theta} and 
\eqn{Xi-evenL-theta} are related by a simple application of Watson's 
identity \eqn{Watson}.

It is rather intriguing that the grand canonical partition function ends up also as a product of 
Jacobi theta functions, similar to 
the superconformal indices of the free hypermultiplets and vector multiplets. The reason 
for this is not clear to us, but it is a manifestation of the modular properties of the Schur index, 
discussed in \cite{Razamat2012}. The same can be said for the expressions we find for 
finite $N$ in Section~\ref{sec:finite}.

In Section~\ref{sec:LargeN} we compute the $d_j$'s at leading order in the large $\mu$ 
expansion, from which we obtain the leading large $N$ contribution to the index. 
In Section~\ref{sec:exactlargeN} 
we focus on the cases of $L=1$ and $L=2$, for which the numerator of
\eqn{one-term} is quadratic and so can be easily factored algebraically, 
and the roots obtained exactly.%
\footnote{Note that the numerator of \eqn{one-term} can also be factored algebraically for 
$L=4$, but we haven't investigated this case.}
This allows us to go beyond the large $N$ limit, and obtain an exact all order expression for 
the index.

\section{Large $N$ limit of the index}
\label{sec:LargeN}

In this section we compute the Schur index for $SU(N)$ circular quivers with $L$ nodes 
in the large $N$ limit and with the product of flavour fugacities set to $1$, so that $U=0$. 
The result for all the theories scales as $N^0$, is independent of the flavour fugacities, and there 
are no perturbative $1/N$ corrections. We address the exponential corrections in 
$N$ for $L=1$ and $L=2$ in the next section. 
In the subsections below we present two different methods which both give the 
same perturbatively exact large $N$ result.

\subsection{Asymptotics from the grand canonical partition function}
\label{polynomialasymptotics}

The first method relies on the expression \eqref{Xi-L-theta} for 
the grand canonical partition function $\Xi_{\vec{n}}$ in terms of the roots of 
a degree $2L$ polynomial. We solve for the roots of the polynomial at large 
$\mu$, from that obtain $\Xi_{\vec{n}}$ and through \eqref{invertXin} find 
$Z_{\vec{n}}$. Taking the sum over the Fourier modes \eqref{ISUN}
and including the prefactors in \eqn{jacobiindexsun}, 
we finally obtain the index up to non-perturbative corrections in the large $N$ limit.

We first compute the large $\mu$ expansion of the $d_j$, 
introduced in \eqref{defdi}. Recall that $X_j=-e^{-2 i d_j}$ are the roots 
of the polynomial
\beq
\label{polynomialdi}
\prod_{a=1}^L \Big(1 + q^{-2 n^{(a)}} X^2 \Big)+ \kappa q^{- n} X^L \,,
\eeq
$X_j$ can be expanded at large $\kappa$ as 
\beq
\label{rootsexpansion}
X_j = X^{(0)}_j \kappa^{\gamma_j}(1 + \cO(\kappa^{\beta_j}))\,,
\eeq
where $X^{(0)}_j$ is a (non zero) constant and $\beta_j< 0$. 
Plugging this ansatz into \eqn{polynomialdi}, and expanding at leading order in $\kappa$, 
the roots must satisfy 
\beq
\begin{cases}
0= q^{- 2n} (X^{(0)}_j )^{2L} \kappa^{2\gamma_j L} + q^{- n} 
\kappa^{\gamma_j L+1} (X^{(0)}_j )^{L} + \cO(\kappa^{2 \gamma_j(L -1)})\,, &\gamma_j>0,
\\
0 = (X^{(0)}_j)^L q^{-n} \kappa + \cO(\kappa^0)\,, & \,\gamma_j =0 \,, \\
0 =1+ \kappa^{1- \gamma_j L} q^{- n} (X^{(0)}_{j})^L + \cO(\kappa^{-2 \gamma_j})
\,, &\gamma_j<0\,.
\end{cases}
\eeq
The second line has no solutions, while the first and third lines each admit 
$L$ solutions with $\gamma_j = \frac{1}{L} $ and $\gamma_j = -\frac{1}{L} $ 
respectively, and with
\beq
X^{(0)}_{j} = e^{\frac{i\pi(2 j + 1)}{L}} q^{\frac{n}{L}} \,,
\qquad
j = 1, \cdots, L\,. 
\eeq

Going to the next order, we find 
that for all of the roots \eqn{rootsexpansion} $\beta_j = - \frac{2}{L}$.
We can then readily deduce the large $\mu$ expansions for $d_j$
\beq
d_{j, \pm}
= \pm \frac{i\mu}{2L} 
+ \frac{(L+1 -2j)\pi}{2L} + \frac{n \pi\tau}{2L}
+ \mathcal{O}\big(\kappa^{-\frac{2}{L}}\big)\,, \qquad j = 1, \cdots, L \,,
\eeq
where the indices differ slightly from the ones used in \eqref{defdi}. 
Using the above expression, we can in turn expand \eqref{Xi-L-theta} 
in the large $\mu$ limit as
\bal
\Xi_{\vec{n}}
=\frac{q^{\sum_{a=1}^L(n^{(a)})^2}}{\vartheta_4^L\vartheta_3^L}
&\prod_{j=1}^{L} 
\vartheta_3\Big(\frac{i\mu}{2 L} + \frac{(L+1 - 2 j)\pi}
{2L} + \frac{n \pi\tau}{2L},q^{\frac{1}{2}}\Big)
\\
&\ {}\times
\vartheta_3\Big(\frac{i\mu}{2 L} + \frac{(L+1 - 2 j)\pi}
{2L} - \frac{n \pi\tau}{2L},q^{\frac{1}{2}}\Big)+ \mathcal{O}(\kappa^{-2/L})\,.
\eal

This last expression involves the product of theta functions shifted by fractions 
of $\pi$. This product can be done using then the identity \eqref{productformula} 
proven in Appendix~\ref{identities}
\bal
\Xi_{\vec{n}} 
&=
\frac{q^{\sum_{a=1}^L(n^{(a)})^2}}{\vartheta_4^L\vartheta_3^L}
\frac{\eta^{2L}(\frac{\tau}{2})}{\eta^2(\frac{L\tau}{2})}
\vartheta_3 \Big(\frac{i\mu}{2} + \frac{n \pi\tau}{2}, q^\frac{L}{2} \Big)
\vartheta_3 \Big(\frac{i\mu}{2} - \frac{n \pi\tau}{2}, q^\frac{L}{2} \Big)
+ \mathcal{O}(\kappa^{-2/L})\,.
\eal
Now using Watson's identities \eqref{Watson}, as well as \eqref{reductionbis}, gives
\begin{align}
\label{sameXin}
\Xi_{\vec{n}} 
&=
q^{\sum_{a=1}^L(n^{(a)})^2} 
\frac{\eta^L(\tau)}{\vartheta_3^L \vartheta_4(0, q^L)\eta(L \tau)}
\left(
\vartheta_3(i\mu, q^L)\vartheta_3(n \pi\tau, q^L)
+\vartheta_2(i\mu, q^L)\vartheta_2(n \pi\tau, q^L)
\right)
\nonumber\\
& \quad 
+ \mathcal{O}(\kappa^{-2/L})\,.
\end{align}

From this expression one can obtain $Z_{\vec{n}}$ \eqref{Inasun} 
via the integral transform \eqref{invertXin} and the expressions for the 
Fourier coefficients of the theta function \eqref{integrals}
\bal
\label{SUNZna}
Z_{\vec{n}}(N)&= 
q^{\sum_{a=1}^L(n^{(a)})^2}
q^{\frac{LN^2}{4}} 
\frac{\eta^L(\tau)}
{2 \vartheta_3^{L} \vartheta_4(0, q^L)\eta(L\tau)} 
\\
& \quad \times 
\left((1+(-1)^N)\vartheta_3(\pi\tau n, q^{L})
+(1-(-1)^N)\vartheta_2(\pi\tau n, q^{L})\right)
+ \dots
\eal

To get the index we need to sum over the Fourier modes $\vec n$, 
as in \eqref{ISUN} 
(recall that $n^{(L)}=0$ \eqn{SUNZl}). 
Using the series 
representation \eqref{thetaseriesproduct} of the theta functions above, we find
\begin{align}
\sum_{\vec{n} \in \bZ^{L-1}} & q^{\sum_{a=1}^{L-1}(n^{(a)})^2} 
e^{-2i N \sum_{a=1}^{L-1} n^{(a)} u^{(a)}}
\vartheta_3(\pi\tau n, q^{L}) 
\nn\\
\label{thetaL}
&= \sum_{\{\vec{n},p \} \in \bZ^{L}} 
q^{\sum_{a=1}^{L-1}(n^{(a)})^2 +L p^2} 
e^{-2i N \sum_{a=1}^{L-1} n^{(a)} u^{(a)}}\\
&
= \sum_{\{\vec{n},p \} \in \bZ^{L}} 
q^{\sum_{a=1}^{L-1}(n^{(a)} + p)^2 + p^2} 
e^{-2i N \sum_{a=1}^{L-1} (n^{(a)}+p) u^{(a)}-2iNp u^{(L)}}
= \prod_{a=1}^{L}\vartheta_3(N u^{(a)}) \,.
\nn
\end{align}
Similarly we obtain
\beq
\sum_{\vec{n} \in \bZ^{L-1}} q^{\sum_{a=1}^{L-1}(n^{(a)})^2} 
e^{-2i N \sum_{a=1}^{L-1} n^{(a)} u^{(a)}}
\vartheta_2(\pi\tau n, q^{L})
= \prod_{a=1}^L\vartheta_2( N u^{(a)}) \,.
\eeq
The sum over $Z_{\vec n}$ is now simple, with only some care required to account 
for the $(-1)^N$ factors. For this, we use the formula 
(see \eqn{quasiperiodicity})
\beq
\label{evenoddtheta}
q^{\frac{N^2}{4}}\vartheta_3 \Big( Nu^{(a)} + \frac{\pi \tau}{2} N\Big)
= 
\begin{cases} 
q^{- i N^2 u^{(a)}} \vartheta_3 ( N u^{(a)} )\,, &N \,\text{even},
\\
q^{- i N^2 u^{(a)}} \vartheta_2 ( N u^{(a)} ) \,, & N\,\text{odd}, 
\end{cases}
\eeq
which gives
\beq
\label{SUNZN}
Z(N,L)=q^{\frac{LN^2}{2}}
\frac{\eta^L(\tau)}{\vartheta_4(0, q^L)\eta(L\tau)}
\prod_{a=1}^L \frac{\vartheta_3 \left( N u^{(a)} + \frac{\pi \tau}{2} N\right)}{\vartheta_3}
+ \dots
\eeq
Substituting this result in \eqn{jacobiindexsun}, we finally obtain
\beq
\label{finallargeNresult}
\mathcal{I} (N,L) = \frac{q^{\frac{L}{6}}} 
{\vartheta_4 (0, q^L )\eta^{L}(\tau)\eta(L\tau)} + \mathcal{O} \big( e^{-c N} \big)\,, 
\qquad c>0\,.
\eeq
Writing the remaining theta function in terms of eta functions this can also be written as
\beq
\label{finallargeNresult2}
\mathcal{I} (N,L)  = \frac{q^{\frac{L}{6}}} 
{\eta^{L}(\tau)\eta^2(\tfrac{L\tau}{2})} + \mathcal{O} \big( e^{-c N} \big)\,.
\eeq
As previously mentioned, for the case of $L=1$ the result is slightly modified, because 
the matter multiplet is in the adjoint rather than bi-fundamental representation 
(see the comment after equation \eqn{za}). 
\eqn{finallargeNresult2} is the main result of this section, 
which we reproduce in the next subsection using different techniques. We find that 
the full perturbative dependence on $N$ is given by this constant term with no subleading 
$1/N$ corrections (ignoring non-perturbative corrections), and that the results 
does not depend on the flavour fugacities. 

To get to the final result we have first integrated over $\mu$ and then summed over 
$\vec n$. For completeness we do it also in the reverse order, first 
summing $\Xi_{\vec{n}}$ over the Fourier modes. This suggests to define an overall 
$\Xi(\kappa)$ as
\bal
\label{grandindex}
\Xi(\kappa, L)&\equiv1+ \sum_{\vec{n} \in \bZ^{L-1}} e^{-2 i N \sum_{a=1}^{L-1} u^{(a)} n^{(a)}}
\sum_{N=1}^\infty Z_{\vec{n}}(N)\kappa^N
\\
&=1+\sum_{\vec{n} \in \bZ^{L-1}}e^{-2 i N \sum_{a=1}^{L-1} u^{(a)} n^{(a)}} (\Xi_{\vec{n}}(\kappa) -1)\,.
\eal
Using \eqref{sameXin} and \eqn{thetaL} we obtain
\beq
\Xi(\kappa, L)
=\frac{\eta^L(\tau)}{\eta^2(\frac{L \tau}{2})}
\left(\vartheta_3(i\mu, q^L)\prod_{a=1}^L \frac{\vartheta_3( Nu^{(a)} )}{\vartheta_3}
+\vartheta_2(i\mu, q^L)\prod_{a=1}^L \frac{\vartheta_2( Nu^{(a)} )}{\vartheta_3}\right)
+ \dots
\eeq

Furthermore, as in the case of $\cN=4$ SYM in \cite{Bourdier:2015wda}, we can define the odd 
and even parts of $\Xi$ as
\beq
\Xi_\pm(\kappa, L)=\frac{1}{2}\left(\Xi(\kappa, L)\pm\Xi(-\kappa, L)\right).
\eeq
Since $\vartheta_3$ is periodic in $\pi$ and $\vartheta_2$ antiperiodic, we find
\bal
\Xi_+(\kappa, L)
&=\frac{\eta^L(\tau)}{\eta^2(\frac{L \tau}{2})}
\vartheta_3(i\mu, q^L)
\prod_{a=1}^L \frac{\vartheta_3( Nu^{(a)} )}{\vartheta_3}
+ \dots
\\
\Xi_-(\kappa, L )
&=\frac{\eta^L(\tau)}{\eta^2(\frac{L \tau}{2})}
\vartheta_2(i\mu, q^L)
\prod_{a=1}^L \frac{\vartheta_2( Nu^{(a)} )}{\vartheta_3}
+ \dots
\eal
Recall the factor in \eqn{jacobiindexsun} relating the index with the rescaled index $Z(N)$
\beq
\frac{q^{- \frac{N^2 L}{2}}q^{\frac{L}{6}}q^{i N^2 \sum_{a} u^{(a)}}}
{\eta^{2L}(\tau)}
\prod_{a=1}^L
\frac{\vartheta_3}{\vartheta_3 \big(N u^{(a)}+ N\frac{\pi\tau}{2} \big)}\,,
\eeq
which due to \eqref{evenoddtheta} 
has a nice alternating behavior between even and odd $N$ 
apart for a factor of $q^{-\frac{N^2L}{4}}$. 
This suggest that we can also define a {\em grand index} as
\beq
\label{grand}
\hat\Xi(\kappa, L )
=1+\sum_{N=1}^\infty q^{\frac{L N^2}{4}}\cI(N)\kappa^N\,.
\eeq
This does not involve all the rescaling factors in
\eqn{jacobiindexsun}, 
and the difference between even and odd $N$ is captured by different 
rescalings of the $\Xi_\pm$ defined above as
\bal
\hat\Xi(\kappa, L )
&=
\Xi_+(\kappa, L )\,
\frac{q^{\frac{L}{6}}}{\eta^{2L}(\tau)}
\prod_{a=1}^L \frac{\vartheta_3}{\vartheta_3( Nu^{(a)} )}
+\Xi_-(\kappa, L )\,
\frac{q^{\frac{L}{6}}}{\eta^{2L}(\tau)}
\prod_{a=1}^L \frac{\vartheta_3}{\vartheta_2( Nu^{(a)} )}
\\&=\frac{q^{\frac{L}{6}}}{\eta^L(\tau)\eta^2(\frac{L \tau}{2})}
\left(\vartheta_3(i\mu, q^L)
+\vartheta_2(i\mu, q^L)\right)
+ \dots
\eal
Equation \eqn{finallargeNresult2} is easily reproduced from the inverse of 
\eqn{grand}, {\it i.e.}, the Fourier expansion of $\hat\Xi$.

\subsection{Asymptotics from the grand potential}
\label{largeNcomplexanalysis}

In the previous subsection we used the formula for $\Xi_{\vec{n}}$ 
in terms of the roots of a polynomial \eqref{Xi-L-theta} and used the large 
$\mu$ expansion of the roots to find 
\eqref{sameXin}, from which we deduced the perturbative part of the large 
$N$ behavior of the index.

We now present an alternative way of obtaining the large $\mu$ limit of 
$\Xi_{\vec n}$ \eqref{sameXin} in the case with vanishing flavour fugacities, 
by applying the large $\mu$ approximation to 
the grand potential \eqref{Jmu}. An analog method was used in the case of 
3-dimensional theories and is instructive as it does not rely on the exact expression 
for $Z_{\vec n;\ell}$, which may not be available in other settings.

To find the grand potential at large $\mu$ we only need the asymptotic behavior 
of $Z_{\vec n;\ell}$ at large $\ell$. Following \cite{Hatsuda:2015oaa},
we use the Mellin-Barnes representation
\beq
\label{Jmucontour} 
J_{\vec{n}}(\mu)
= -\int_{c- i\infty}^{c+i\infty} \frac{d \ell}{2\pi i} \,
\frac{\pi}{\sin\pi\ell}\frac{{Z}_{\vec{n};\ell}}{\ell} e^{\ell \mu}\,,
\quad
0 < c <1\,,
\eeq
and extract the leading order in the large $\mu$ from the 
poles of $Z_{\vec{n};\ell}$ with largest $\text{Re}(\ell)$.

The representation \eqn{Jmucontour} requires some explanation and justification. 
We first write $Z_{\vec{n};\ell}$ \eqref{SUNZl} as an analytic function of $\ell$ by 
splitting it into two sums, 
one for positive $p$ and one for strictly negative $p$. 
Denoting the sum over the terms with positive $p$ as $Z^+_{\vec{n};\ell}$, we have
\bal
Z^+_{\vec{n};\ell}&= 
\sum_{p=0}^\infty
\frac{q^{\ell \sum_{a=1}^L(p-n^{(a)} +\half)}}
{\prod_{a=1}^L \big(1 + q^{2(p- n^{(a)} +\half)}\big)^\ell} 
\\ &=
q^{-\ell \sum_{a=1}^L( n^{(a)} -\half)} \sum_{p=0}^\infty
q^{\ell L p} \sum_{\vec{k} \in \bZ^L_+} 
q^{2 \sum_{a=1}^L k^{(a)} (p-n^{(a)} +\half)} \prod_{a=1}^L \binom{-\ell}{k^{(a)}} 
\\&= 
q^{-\ell \sum_{a=1}^L(n^{(a)} -\half)} 
\sum_{\vec{k} \in \bZ^L_+} q^{-2 \sum_{a=1}^L k^{(a)}(n^{(a)} -\half)} 
\prod_{a=1}^L \binom{-\ell}{k^{(a)}} \sum_{p=0}^\infty
q^{p(\ell L + 2 \sum_{a=1}^L k^{(a)})} \,.
\eal
Doing the summation over $p$, we obtain
\beq
\label{Zln}
Z^+_{\vec{n};\ell}
= \sum_{\vec{k} \in \bZ^L_+} \frac{q^{-\sum_{a=1}^L(2 k^{(a)}+ \ell)
(n^{(a)} -\half)}}
{1 - q^{(\ell L + 2 \sum_{a=1}^L k^{(a)})}}\prod_{a=1}^L \binom{-\ell}{k^{(a)}}\,.
\eeq
This final form admits an analytical continuation in $\ell$ to the complex plane, 
and a similar argument can be used for $Z^{-}_{\vec{n};\ell}$, 
which is obtained by replacing $n^{(a)} \rightarrow - n^{(a)}$.
For negative values of $\mu$ one can then 
compute the r.h.s. of \eqref{Jmucontour} by closing the contour with 
an infinite half circle enclosing the simple poles due 
to 
$\pi/\sin\pi\ell$ 
at positive values of $\ell$, but none of the poles due to $Z_{\vec{n};\ell}$. 
Using the fact that
\beq
\mathop{\Res}_{\ell=n}
\frac{\pi}{\sin\pi\ell}=(-1)^n\,,
\eeq
and the fact that the evaluation of the integral on the remaining part of the contour gives zero, 
we recover the representation \eqn{Jmu} as an infinite sum, which is indeed 
convergent for negative $\mu$.

To analytically continue $J_{\vec{n}}(\mu)$ to positive values of $\mu$, we close 
the contour in \eqref{Jmucontour} with an infinite half-circle in the 
$\text{Re}(\ell)\leq c$ half-plane. 
In this enclosed region, the poles of $Z_{\vec{n}; \ell}$ 
and of the cosecant 
are then at
\bal
\label{poles1}
\ell&=- \frac{2}{L} \sum_{a=1}^L k^{(a)} + \frac{2 l}{L\tau} \,, 
&&\qquad
k^{(a)} \in \mathbb{N} \,, 
\quad 
l \in \bZ\,,
\\
\ell
&
= -n\,,
&&\qquad
n \in \mathbb{N}\,. 
\eal
It can be shown that the contour integrals coming 
from the integration over the infinite half-circle do not contribute, so that \eqref{Jmucontour} 
is determined only by the residues of the poles \eqref{poles1}.

As explained in the previous section
we are ultimately interested in $J_{\vec{n}}(\mu)$ for large $\mu$. The poles  
that are not on the imaginary axis are exponentially 
suppressed in this limit. We can thus write
\beq
\label{Jmures}
J_{\vec{n}}(\mu)= - \sum_{m \in \bZ} 
\mathop{\Res}_{\ell=\frac{2 m}{L\tau}} 
\frac{\pi}{\sin\pi\ell}\frac{Z_{\vec{n};\ell}}{\ell}
e^{\ell \mu} +\cO(e^{-\frac{2\mu}{L}})\,,
\eeq
where the scaling in $\mu$ of the next to leading order can be deduced 
from the lattices \eqref{poles1}. For 
the residue of the pole at $\ell=0$, we obtain
\bal
&-\mathop{\Res}_{\ell=0}
\frac{\pi}{\sin\pi\ell}\frac{Z_{\vec{n};\ell}}{\ell}
e^{\ell \mu} 
\\&= i\frac{4 \pi^2 +L^2 \pi^2 \tau^2 + 12 \mu^2 - 12 \pi^2 \tau^2 n^2}{12 L \pi\tau} 
+ \sum_{a= 1}^L \sum_{k =1}^\infty \cosh 2 i \pi\tau k n^{(a)} 
\frac{(-1)^{k}}{k \sinh - i\pi\tau k}
\\&
=i\frac{4 \pi^2 +L^2 \pi^2 \tau^2 + 12 \mu^2 - 12 \pi^2 \tau^2 n^2}{12 L \pi\tau} 
-\sum_{a=1}^L\left( \log \frac{\vartheta_3(\tau \pi n^{(a)})}{\vartheta_3}
+\frac{i\pi\tau}{12}+\frac{1}{6} \log \frac{4}{k k'}\right),
\eal
where the sum over $k$ was done using \eqref{cossinhseries}.

The sum over the poles on the imaginary axis but away from the origin gives
\bal
\label{otherpoles}
&-\sum_{m\neq0}\mathop{\Res}_{\ell=\frac{2 m}{L\tau}} 
\frac{\pi}{\sin\pi\ell}\frac{Z_{\vec{n};\ell}}{\ell}
e^{\ell \mu} 
= \sum_{m\neq0}(-1)^{m+1} \frac{e^{\frac{2 m \mu}{L\tau}} 
\cos\frac{2\pi m n}{L}}{m \sinh \frac{2 i\pi m}{L\tau}}
\\&\qquad
= \log \frac{\vartheta_3 \big(\frac{\pi}{L}\big(\frac{i\mu}{\pi\tau} + n \big), 
{q'}^{\frac{2}{L}} \big)
\vartheta_3 \big(\frac{\pi}{L}\big(\frac{i\mu}{\pi\tau} - n \big), 
{q'}^{\frac{2}{L}} \big)}
{\vartheta_3^2 \big(0, {q'}^{\frac{2}{L}} \big)} 
- \frac{i\pi}{3 L\tau} + \frac{1}{3} \log \frac{4}{\tilde{k} \tilde{k}'} \,.
\eal
This sum was again done using \eqref{cossinhseries} but with the complement nome 
and corresponding modulus
\beq
q' = e^{- \frac{i\pi}{\tau}} \,,
\quad
\tilde{k} = \frac{\vartheta_2^2 
(0, {q'}^{\frac{2}{L}})}{\vartheta_3^2(0, {q'}^{\frac{2}{L}})} \,, 
\quad
{\tilde{k'}} = 
\frac{\vartheta_4^2(0, {q'}^{\frac{2}{L}})}{\vartheta_3^2(0, {q'}^{\frac{2}{L}})}
\,.
\eeq
Applying a modular transformation \eqn{Jacobiimaginary} to \eqn{otherpoles} gives
\beq
\log\frac{\vartheta_3 \big(\frac{i\mu}{2} + \frac{\pi\tau n}{2},{q}^{\frac{L}{2}} \big)
\vartheta_3 \big(\frac{i\mu}{2} - \frac{\pi\tau n}{2}, {q}^{\frac{L}{2}}\big)}
{\vartheta_3^2 \big(0, q^{\frac{L}{2}} \big)}
- \frac{i\mu^2}{\pi\tau L} + \frac{i\pi\tau n^2}{L}
- \frac{i\pi}{3 L\tau} + \frac{1}{3} \log \frac{4}{\tilde{k} \tilde{k}'} \,.
\eeq

Putting the contributions from all the poles on the imaginary axis together, we obtain
\bal
\label{sumofresiduessun}
J_{\vec{n}}(\mu)
&= \log\frac{\vartheta_3 \big(\frac{i\mu}{2} + \frac{\pi\tau n}{2},{q}^{\frac{L}{2}} \big)
\vartheta_3 \big(\frac{i\mu}{2} - \frac{\pi\tau n}{2}, {q}^{\frac{L}{2}}\big)}
{\vartheta_3^2 \big(0, q^{\frac{L}{2}} \big)}
\\
& \quad 
+ \frac{2-L}{3} \log 2+ \frac{1}{3} \log \frac{(k k')^{L/2}}{\tilde{k} \tilde{k}'} 
- \sum_{a=1}^L \log \frac{\vartheta_3(\tau \pi n^{(a)})}
{\vartheta_3}
+ \cO(e^{-2\mu /L})\,.
\eal 
Finally we use Watson's identity \eqref{Watson} to rewrite the product of theta functions in the 
first line in terms of theta functions with nome $q^L$. We also replace $k,k',\tilde{k},\tilde{k}'$, applying modular transformations to the latter two.
Then we use the quasi-periodicity of the theta function \eqref{quasiperiodicity}, 
and use \eqref{reduction} and \eqref{reductionbis} 
to express the result in terms of the Dedekind eta functions to find
\bal
J_{\vec{n}}(\mu)&= \log
\left(\vartheta_3(i\mu, q^{L})\vartheta_3(\pi\tau n, q^{L})
+\vartheta_2(i\mu, q^{L})\vartheta_2(\pi\tau n, q^{L})\right)
\\
& \quad 
+ \log \frac{\eta(\tau)^L}
{\eta(L\tau)\vartheta_3^{L} \vartheta_4(0, q^L)} 
+ i\pi\tau \sum_{a=1}^L (n^{(a)} )^2 + \cO (e^{-2 \mu / L} )\,.
\eal
We finally obtain
\bal
\label{mudependentsunJ}
\Xi_{\vec{n}}(\kappa)&= 
q^{\sum_{a=1}^L (n^{(a)} )^2} 
\frac{\eta(\tau)^L}
{\vartheta_3^{L} \vartheta_4 (0, q^L )\eta(L\tau)} 
\\
& \quad \times 
\left(\vartheta_3 (i\mu, q^{L})\vartheta_3 (\pi\tau n, q^{L})
+\vartheta_2 (i\mu, q^{L})\vartheta_2 (\pi\tau n, q^{L})\right)
+ \cO (\kappa^{-2/L})\,,
\eal
which is identical to \eqref{sameXin}.

\section{Exact large $N$ expansions for short quivers}
\label{sec:exactlargeN}

For quivers with one or two nodes we can compute the Schur index exactly, 
without having to resort to perturbative techniques. Recall that the grand 
partition function can be expressed by a product of theta functions 
\eqn{Xi-L-theta} evaluated at the roots of the polynomial \eqn{one-term}
\beq
\label{polynomial}
X^L q^{- n}\,\kappa 
+ 
\prod_{a=1}^L\big(1+X^2q^{-2n^{(a)}}\big)\,.
\eeq
For $L=1$ and $L=2$ this polynomial is quadratic in $X$ and $X^2$ 
respectively, and so the roots are simply algebraic.%
\footnote{\label{L=4}For the theory with $L=4$ the polynomial is quartic 
and so can also be factored algebraically. It would 
be interesting to see if a similar analysis would also give a complete solution for 
the index of this theory.} 
This results in completely explicit expression for the grand partition functions which 
allow us to find closed form expressions for the indices of these theories.
We start by reviewing, the $L=1$ calculation carried out in \cite{Bourdier:2015wda},
and then show that the same discussion can be applied to the $L=2$ case.

\subsection{Single node, $\cN=4$ SYM}
\label{sec:N4}

For the single node theory 
(with the only flavour fugacity $e^{2 iu^{(1)}} = e^{2 i U}$ set to one), 
the polynomial \eqn{polynomial} can be factored as
\beq
\label{factor1}
X\kappa+1+X^2
=\left(1+\tfrac{\kappa+\sqrt{\kappa^2-4}}{2}\,X\right)
\left(1+\tfrac{\kappa-\sqrt{\kappa^2-4}}{2}\,X\right), 
\eeq
Comparing with \eqn{defdi} we readily obtain
$d_\pm =\pm\frac{i}{2}\log\frac{\kappa+\sqrt{\kappa^2-4}}{2}
=\pm\frac{1}{2}\arccos\frac{\kappa}{2}$. 

Unlike the cases of $L>1$, there are no Fourier modes to sum over, giving 
a single free Fermi gas whose grand partition function is%
\footnote{Although the matrix model 
still has a delta function coming from the tracelessness condition of $SU(N)$,
the Kronecker delta in \eqref{SUNZl} ensures that only the mode with $n =0$ 
contributes.}
\eqn{Xi-L-theta}
\beq
\label{Xi-N4}
\Xi(\kappa, 1)
=\frac{\vartheta_3^2\big(\frac{1}{2}
\arccos\frac{\kappa}{2},q^{\frac{1}{2}}\big)}{\vartheta_3\vartheta_4}
=\frac{1}{\vartheta_4}
\left[\vartheta_3\Big(\arccos \frac{\kappa}{2}\Big)
+\frac{\vartheta_2}{\vartheta_3}\,\vartheta_2 \Big(\arccos \frac{\kappa}{2}\Big)\right].
\eeq
This is indeed the expression found in \cite{Bourdier:2015wda}. 
Recall that in terms of the grand partition function, the index is given by%
\footnote{\label{adjoint2}%
The prime indicates that the hypermultiplet is in the adjoint rather 
than the bi-fundamental representation of $SU(N)$. 
The difference is the removal of one free adjoint hypermultiplet, which 
introduces an additional factor of 
$ q^\frac{1}{12} \eta^2(\frac{\tau}{2})\eta^{-2}(\tau)$ 
in \eqn{N=4indexfromgrand} compared with \eqn{jacobiindexsun} 
with vanishing $u$'s and $L=1$.}
\beq
\label{N=4indexfromgrand}
\mathcal{I}\new{'} (N,1)= 
\frac{q^{-\frac{N^2}{2}} 
q^{\frac{1}{4}}\eta^2(\frac{\tau}{2}) \vartheta_3}
{\eta^4(\tau)\vartheta_3 ( \frac{\pi \tau}{2} N)} 
\int_{- i\pi}^{i\pi} \frac{d \mu}{2\pi i}\,e^{-\mu N} \Xi^{\cN = 4}(e^\mu)\,. 
\eeq
In \cite{Bourdier:2015wda} the integral over $\mu$ was evaluated by studying the large 
$\mu$ expansion of the integrand. We proceed here in a slightly different way, performing 
instead the complete expansion in powers of $e^\mu$ and $q$. Since the calculation is 
ultimately exact, we arrive at the same result.

Expanding the square of the theta function in the middle expression of \eqn{Xi-N4} gives 
\beq
e^{-\mu N} \Xi (e^\mu, 1 )= \frac{e^{-\mu N}}{\vartheta_3 \vartheta_4} 
\sum_{m=-\infty}^\infty \sum_{j= - \infty}^\infty 
q^{\frac{1}{2}(m^2 + j^2)} \left(\tfrac{e^\mu +\sqrt{e^{2\mu}-4}}{2}\right)^{m+j}\,.
\eeq
Applying the expansion formula \eqn{expansion1} this is
\beq
\frac{1}{\vartheta_3 \vartheta_4} \sum_{m=-\infty}^\infty 
\sum_{j= -\infty}^\infty \sum_{s=0}^\infty
q^{\frac{1}{2}(m^2 + j^2)} e^{\mu(m+j-2s-N)} 
\frac{(-1)^s(m+j)(m+j-s-1)!}{s!(m+j-2s)!}
\eeq
Integrating over $\mu$ simply gives a Kronecker delta $\delta_{m+j-2s-N} $, 
which removes the sum over $m$ 
\bal
Z(N, 1)
&=\int_{- i\pi}^{i\pi} \frac{d \mu}{2 \pi i}e^{-\mu N} \Xi^{\cN = 4}(e^\mu)
\\
&
= \frac{1}{\vartheta_3 \vartheta_4}\sum_{j= -\infty}^\infty 
\sum_{s=0}^\infty
q^{(j -s -\frac{N}{2})^2 +(s+\frac{N}{2})^2} \frac{(-1)^s(N+2s)(N+s-1)!}{s!N!}\,.
\eal
Finally evaluating the sum over $j$ and including the prefactors from \eqn{N=4indexfromgrand} yields
\bal
\label{largeN}
\mathcal{I}' (N,1)
&= \frac{q^{\frac{1}{4}}}{\eta^3(\tau)} \sum_{s=0}^\infty
(-1)^s(N+2s)\frac{(N+s-1)!}{s!N!}q^{N s +s^2}\\
&=\frac{q^{\frac{1}{4}}}{\eta^3(\tau)}
\sum_{s=0}^\infty(-1)^s \left[\binom{N+s}{N}+\binom{N+s-1}{N}\right]
q^{N s +s^2}\,.
\eal

At leading order at large $N$ this is simply $q^{1/4}/\eta^2(\tau)$, which differs from 
\eqn{finallargeNresult2} by the contribution of one free hypermultiplet 
(see the comment after equation \eqn{za}). 
As discussed in \cite{Bourdier:2015wda} it would be interesting to find 
an interpretation for the exponential corrections, possibly as D3-brane giant gravitons in 
$AdS_5\times S^5$.

\subsection{Two nodes}
\label{sec:L2}

For the two-node quiver, $L=2$, 
the polynomial \eqn{polynomial} can be factored as
\beq
\label{factor2}
\kappa q^{-n} X^2+(1+ X^2)(1+q^{-2n}X^2)
= \big(1+\tfrac{\tilde\kappa +\sqrt{\tilde\kappa^2-4}}{2}\,q^{-n}X^2\big)
\big(1+\tfrac{\tilde\kappa -\sqrt{\tilde\kappa^2-4}}{2}\,q^{-n}X^2 \big),
\eeq
where $ \tilde\kappa = \kappa + q^n + q^{-n} $. Comparing with \eqn{defditilde} we obtain 
$\tilde d_{\pm} =-\frac{\pi\tau n}{2}\pm\frac{1}{2}
\arccos\frac{\tilde\kappa}{2} $ and substituting into \eqn{Xi-evenL-theta} gives
\beq
\label{Xi-2}
\Xi_{n}(\kappa,2)
=\frac{q^{n^2}}{\vartheta_3^2}
\vartheta_3\Big(\frac{\pi\tau n}{2}+\frac{1}{2}\arccos\frac{\tilde\kappa}{2}\Big)
\vartheta_3\Big(\frac{\pi\tau n}{2}-\frac{1}{2}\arccos\frac{\tilde\kappa}{2}\Big).
\eeq
Using Watson's identity \eqn{Watson}, this can also be written as the sum of theta 
functions with nome $q^2$, ({\it c.f.}, the last expression in \eqn{Xi-N4}), but this 
representation will not be simpler for us.

In terms of the grand partition function, the index with flavour fugacities such that 
$ u \equiv u^{(1)} = - u^{(2)}$, is given by
(see \eqn{jacobiindexsun}, \eqn{ISUN} and \eqn{invertXin})
\beq
\label{2nodesindexfromgrand}
\mathcal{I} (N,2)= 
\frac{q^{-N^2} q^{\frac{1}{3}} 
\vartheta_3^2}{\eta^4(\tau)\prod_{\pm}\vartheta_3( N\frac{\pi \tau}{2} \pm N u)} 
\sum_{n=-\infty}^\infty \frac{e^{-2 i N n u}}{2 \pi i} 
\int_{- i\pi}^{i\pi} d \mu\,e^{-\mu N} \Xi_n^{L=2}(e^\mu)
\,.
\eeq
One could proceed by evaluating the large $\mu$ expansion of the integrand, 
but it turns out to be simpler to perform instead the full expansion of the grand 
partition function in powers of $e^\mu$ and $q$.

Expanding the theta functions in \eqn{Xi-2}, the integrand of
\eqn{2nodesindexfromgrand} can be written as
\beq
e^{-\mu N} \Xi_n  = \frac{e^{-\mu N}}{\vartheta_3^2}
\sum_{m=-\infty}^{\infty}\sum_{j=-\infty}^{\infty}q^{n^2 +m^2 +j^2}
q^{n(m+j)} 
\Big(\tfrac{e^\mu+q^n + q^{-n} +
\sqrt{({e^\mu}+q^n + q^{-n})^2-4}}{2}\Big)^{m-j}\,.
\eeq
Using the expansion formula \eqn{expansionformula} this is
\bal
&\frac{1}{\vartheta_3^2}
\sum_{m=-\infty}^{\infty}\sum_{j=-\infty}^{\infty} 
\sum_{k=0}^{\infty}\sum_{l=0}^{\infty} q^{n^2+m^2 +j^2} 
q^{n(m+j+k-l)} e^{\mu(m-j-k-l-N)} \\
&\hspace{3.8cm}\times(m-j)
\frac{(m-j-k-1)!(m-j-l-1)!}{k!l!(m-j-k-l)!(m-j-k-l-1)!}\,.
\eal
Integrating over $\mu$ gives a Kronecker delta $\delta_{m-j-k-l-N}$, 
which removes the sum over $m$ 
\bal
Z_n=
\int_{-i\pi}^{i\pi} \frac{d \mu}{2 \pi i} \,e^{-\mu N} \Xi_{n} 
&=\frac{1}{\vartheta_3^2}\sum_{j=-\infty}^{\infty}\sum_{k=0}^{\infty}
\sum_{l=0}^{\infty} q^{n^2+(j+k+l+N)^2 +j^2} q^{n(2j+2k+N)} \\
&\hspace{2cm}\times(N+l+k)\frac{(N+k-1)!(N+l-1)!}{N!(N-1)! k!l!}
\,.
\eal
Summing over $n$ 
\eqn{ISUN} 
then gives 
\bal
Z(N,2)
&=\sum_{n=-\infty}^\infty
e^{- 2i N n u}
Z_n
=\frac{\vartheta_3 ( \frac{\pi \tau}{2}N -u N )}{\vartheta_3^2}
\sum_{j=-\infty}^{\infty}\sum_{k=0}^{\infty}
\sum_{l=0}^{\infty} e^{2 i u N(j+k)} \\
& \quad{}\times
q^{(j+k+l+N)^2 +j^2 -(j+k)(j+k+N)} 
(N+l+k)\frac{(N+k-1)!(N+l-1)!}{N!(N-1)!k!l!}\,.
\eal
Finally evaluating the sum over $j$ and including the 
prefactors from \eqn{2nodesindexfromgrand} we obtain
\beq
\label{I2largeN}
\cI (N, 2)= \frac{q^\frac{1}{3}}{\eta^4(\tau)} 
\sum_{k=0}^\infty \sum_{l=0}^\infty(N+k+l)
\frac{(N+k-1)!(N+l-1)!}{N!(N-1)!k! l!} q^{N(k+l)+2 kl}
e^{2i u N (k-l)}\,.
\eeq
Alternatively this can be written as
\bal
\frac{q^\frac{1}{3}}{\eta^4(\tau)} 
&\sum_{k=0}^\infty 
\sum_{l=0}^\infty 
\left[ \binom{N+k}{N} \binom{N+l-1}{N-1}+\binom{N+k -1}{N-1}
\binom{N+l-1}{N}\right] \\
& \hspace{2cm} \times q^{N(k+l)+2 kl} e^{2i u N (k-l)}\,.
\eal
At leading order at large $N$ this is simply 
\beq
\cI(N,2)= \frac{q^\frac{1}{3}}{\eta^4(\tau)} + \cO(q^{N})\,,
\eeq
in agreement with \eqn{finallargeNresult2}. 
Here we see explicitly how the dependence on $u$ appears from terms in the sum 
with $k-l\neq0$, all of which are exponentially suppressed at large $N$.

As in the case of $\cN=4$ SYM in the previous section, 
the large $N$ expansion \eqn{I2largeN} begs for a holographic interpretation 
(at least for $u=0$). 
For $\cN=4$ there is a single sum \eqn{largeN} while here there 
is a double sum. In both cases the leading exponential term is proportional to $N$, suggesting a 
D3-brane interpretation. 
The double sum could correspond to two different types of D3 giant gravitons, 
with the extra $2kl$ term signifying some interaction between the two stacks of branes. It would 
be interesting to find appropriate supergravity solutions and/or brane embeddings that would 
reproduce this structure.

\section{Finite $N$ results for short quivers}
\label{sec:finite}

In \cite{Bourdier:2015wda} the index of the single 
node quiver (without flavour fugacity) was also written 
in closed form for finite values of $N$ in terms of complete elliptic integrals. 
This was done by studying the spectral traces \eqn{SUNZl},
which for $L=1$ are particularly simple
\beq
Z^{\cN=4}_\ell = \sum_{p \in \bZ} \left(\frac{1}{2 \cosh i\pi\tau(p-\half)}\right)^\ell\,.
\eeq
These sums can be performed using the algorithm of \cite{Zucker1979}. 
The result can then be easily recombined 
using \eqn{ZNZl} and \eqn{jacobiindexsun} 
(with the additional factor
in 
footnote~\ref{adjoint2})
to recover the index. For $N=1,\cdots, 4$ the 
results thus obtained are 
\bal
\label{finiteN-N=4}
\mathcal{I}'(1,1)
&= \frac{\eta^2(\frac{\tau}{2})\sqrt{k}}{\eta^4(\tau)} \frac{K}{\pi} \,,
\qquad 
\mathcal{I}'(2,1)
= \frac{q^{-\frac{3}{4}}\eta^2(\frac{\tau}{2})}{\eta^4(\tau)} \frac{-E K +K^2}{2 \pi^2} \,, \\
\mathcal{I}'(3,1)
&=\frac{q^{-2} \eta^2(\frac{\tau}{2})\sqrt{k}}{\eta^4(\tau)} 
\bigg(\frac{-3 E K^2+(2-k^2)K^3}{6 \pi^3}
+ \frac{K}{24 \pi}\bigg)\,,\\
\mathcal{I}'(4,1)
&= \frac{q^{-\frac{15}{4}} \eta^2(\frac{\tau}{2})}{\eta^4(\tau)}
\bigg(\frac{3 E^2 K^2 -6 E K^3 +(3-2k^2)K^2}{24 \pi^4} - \frac{E K - K^2}{24 \pi^2} \bigg)\,,
\eal 
where $K \equiv K(k^2)$ and $E \equiv E(k^2)$ are complete elliptic integrals of the first 
and second kind respectively, with elliptic modulus given by 
$k ={\vartheta_2^2/\vartheta_3^2}$.

For quivers with more than a single node, we find that computing the spectral 
traces becomes intractable, due to the nontrivial dependence of \eqn{SUNZl} on the 
Fourier modes. In the case 
with no flavour fugacities 
we are still able to proceed by a 
number of alternate methods
(which work perfectly well also for the single node case). 
The first two methods apply to the case of $L=2$ and are based on 
the exact solution 
and the large $N$ expansion in 
Section~\ref{sec:exactlargeN}. 
In the next subsection we use the explicit expression for the grand 
partition function expanded at small $\kappa$ to find the 
result for $N=2$. In the following subsection we use the exact large 
$N$ expansion of the index \eqn{I2largeN} and resum it for finite values of $N$. 
Finally we address some 3-node and 4-node quivers by 
guessing a finite basis of polynomials of elliptic integrals 
and fixing the coefficient by comparing their $q$-expansion to the 
representation of the index as the sum \eqn{ISUN}, \eqn{SUNZl}. 
This can in principle be applied to quivers of arbitrary length and 
with arbitrary rank, but requires significant computational 
resources when either becomes large.

\subsection{Expanding the grand partition function}
\label{sec:expandXi}

Recall that the grand partition 
function for $L=2$ is defined as \eqn{grandpartition}
\beq
\Xi_n (\kappa,2) =1+ \sum_{N=1}^\infty Z_n(N, 2)\kappa^N .
\eeq
Since we found the left hand side in closed form \eqn{Xi-2}, 
we can recover $Z^{L=2}_n(N)$ for finite values of $N$. 
The index is then given by the sum \eqn{ISUN} 
together with the prefactors from \eqn{jacobiindexsun},
which in the case without flavour fugacities becomes
\beq
\label{ISUN-L=2}
\cI (N,2)= \frac{q^{-N^2} 
q^{\frac{1}{3}} \vartheta_3^2}{\eta^4(\tau) 
\vartheta_3^2(\frac{N \pi \tau}{2})}
\sum_{n=-\infty}^\infty Z_n(N,2)\,.
\eeq

For instance, the coefficient of $\kappa^2$ gives 
\beq
\label{ZnN=2a}
Z_{n}(2,2)=\frac{q^{n^2}}{32}
\left(\frac{\vartheta_3'' \vartheta_3(n \pi\tau)}{\vartheta_3^2 \sin^2 n \pi\tau} 
+ \frac{\sin n\pi\tau \vartheta_3''(n\pi\tau)- 
2\cos n \pi\tau \vartheta_3'(n \pi\tau)}{\vartheta_3 \sin^3 n \pi\tau} \right).
\eeq
For $n=0$ this is
\begin{align}
\label{ZnN=2}
Z_0(2, 2)
&= \frac{1}{192 \vartheta_3} \Big(4 \vartheta_3 \vartheta_3'' +3 \vartheta_3''^2 
+ \vartheta_3\vartheta_3^{(4)} \Big)
\\&
= \frac{3 E^2 K^2 -6(1-k^2)E K^3 +(1-k^2)(3-2k^2)K^4}{6 \pi^4} 
+ \frac{(1-k^2)K^2 - K E}{12 \pi^2} \,.
\nn
\end{align}
To get the expression in the last line, 
one can apply the heat equation satisfied by all Jacobi theta functions and convert 
the derivatives into $\tau$ derivatives. Then one can apply the standard 
relation $\vartheta_3= \sqrt{\frac{2K}{\pi}}$ together with 
\eqn{tauderivatives} to reduce everything to complete elliptic integrals.

For $n\neq0$ the partition function \eqn{ZnN=2a} is
\beq
\label{ZnN=2b}
Z_{n \neq0}(2, 2)
= \frac{\vartheta_3''}{\vartheta_3} \frac{1}{16 \sin^2 n \pi\tau} 
+ \frac{i n \cos n \pi\tau - n^2 \sin n \pi\tau}{8 \sin^3 n \pi\tau} \,,
\eeq
where we have used
\beq
\vartheta_3(n \pi\tau)= 
q^{-n^2} \vartheta_3 \,,
\qquad
\vartheta_3'(n \pi\tau)= -2 i n q^{-n^2}\vartheta_3 \,,
\qquad
\vartheta_3''(n \pi\tau)= q^{-n^2} 
(\vartheta_3'' - 4 n^2 \vartheta_3)\,.
\eeq
The sum over $n$ of the first term in \eqn{ZnN=2b} has been evaluated in \cite{Zucker1979}
\bal
\sum_{n \neq0}\frac{1}{16\sin^2 n \pi\tau} = 
\frac{K}{12 \pi^2} \left(3 E -(2-k^2)K \right)
-\frac{1}{48} \,,
\eal
and the prefactor can be written in terms of elliptic integrals as%
\bal
\frac{\vartheta_3''}{\vartheta_3} =
\frac{-4 K E+4(1-k^2)K^2}{\pi^2} \,.
\eal
The sum over $n\neq0$ of the second term vanishes since%
\footnote{This equality can be easily verified by studying the $q$ expansions.}
\beq
\label{cancellation} 
\sum_{n\neq0} \frac{i n \cos n \pi\tau}{8 \sin^3 n \pi\tau}
=\sum_{n\neq0}\frac{n^2}{8 \sin^2 n \pi\tau}\,.
\eeq
Putting this together we obtain
\bal
\label{I2-N=2}
\mathcal{I}(2, 2)&= 
\frac{q^{-\frac{5}{3}}}{\eta^4(\tau)} 
\frac{-3 E^2 K^2 + 2(2-k^2)E K^3 -(1-k^2)K^4}{6 \pi^4}\,,
\eal
One can apply this procedure to higher values of $N$, 
but we find the approach of the next subsection 
to be more efficient.

\subsection{Resumming the large $N$ expansion}

Here we take the result of Section~\ref{sec:exactlargeN} for the exact large 
$N$ expansion \eqn{I2largeN} and resum it for finite values of $N$. 
Inspired by the techniques of 
\cite{Zucker1979}, we find a systematic approach 
to computing this double infinite sum.

The details differ slightly for even and odd 
$N$. First we consider \eqn{I2largeN} 
with $u=0$
for even $N=2 r$
\beq
\cI(2 r, 2)= \frac{q^\frac{1}{3}}{\eta^4(\tau)} 
\sum_{k=-r+1}^\infty \sum_{l=-r+1}^\infty(2r+k+l)
\frac{(2r+k-1)!(2r+l-1)!}{(2r)!(2r-1)!k! l!} q^{2r(k+l)+2kl}\,.
\eeq
Notice that compared to \eqn{I2largeN} we have extended the 
sums to include negative values of $k$ and $l$, for which the 
summand clearly vanishes. Applying the formula
\beq
\frac{(2r+k-1)!}{k!} = \sum_{m=0}^{r-1} \alpha_m(r)(r+k)^{2r-2m-1},
\eeq
where $\alpha_{m}(r)$ are numerical coefficients generated by 
\beq
\sum_{m=0}^{r-1}\alpha_m(r)t^m=\prod_{j=1}^{r-1}(1-j^2t)\,,
\eeq
and writing the sums over $k$ and $l$ in terms of indices $j=k+r$, 
$n=l+r$ yields
\bal
\label{I-L=2-Neven}
&\cI(2 r,2)= 
\frac{q^\frac{1}{3}q^{-2 r^2}}{\eta^4(\tau)(2r-1)!(2r)!} 
\sum_{m=0}^{r-1} \sum_{\smash{m'} =0}^{r-1}\alpha_m(r)\alpha_{m'}(r)\\ 
& \hspace{5.5cm} \times \sum_{j=1}^\infty \sum_{n=1}^\infty 
(j+n)j^{2r-2m-1} n^{2r-2m'-1} q^{2 j n} \,.
\eal

We are now faced by (finitely many) double infinite sums of the form
\beq
\sum_{j=1}^\infty \sum_{n=1}^\infty j^a n^{a+2s+1} q^{2 j n}
=\frac{\partial_\tau^a}{(2 \pi i)^a} 
\sum_{j=1}^\infty \sum_{n=1}^\infty n^{2s+1} q^{2 j n}
= \frac{\partial_\tau^a}{(2 \pi i)^a} A_{2s+1} \,, 
\qquad 
a,s\in \mathbb{N}\,,
\eeq
and likewise with $j\leftrightarrow n$. 
The quantities $A_{2s+1} = \sum_{n=1}^\infty n^{2s+1} \frac{q^{2n}}{1-q^{2n}} $ 
played a central role also in the evaluation of certain hyperbolic sums in \cite{Zucker1979}. 
They are be generated by%
\footnote{$\ns$ and $\sn$ used below are standard Jacobi elliptic functions 
($\ns=1/\sn$).}
\beq
\label{Agen}
\sum_{s=0}^\infty(-1)^s A_{2s+1} \frac{(2t)^{2s}}{2s!} 
= \frac{K(K-E)}{4 \pi^2} + \frac{1}{8 \sin^2 t} - \frac{K^2}{4\pi^2} 
\ns^2 \Big(\frac{2 K t}{\pi}, k^2\Big)\,,
\eeq
Arbitrary numbers of $\tau$ derivatives of the $A_{2s+1}$ can be easily 
evaluated by applying the formulas 
\bal
\label{tauderivatives}
\frac{\partial_\tau}{2 \pi i} k &= \frac{k(1-k^2)K^2}{\pi^2} \,,\\
\frac{\partial_\tau}{2 \pi i} K &= \frac{E K^2 -(1-k^2)K^3}{\pi^2}\,,\\
\frac{\partial_\tau}{2 \pi i} E &=\frac{(1-k^2)(EK^2-K^3)}{\pi^2} \,. 
\eal

Let us now turn to the case of odd $N=2r +1$. Analogously to the
even case, the formula 
\beq
\frac{(2r +k)!}{k!} = 2^{-2r} \sum_{m=0}^{r} \tilde\alpha_m(r)
(2r+2k+1)^{2r-2m}\,,
\eeq
where $\tilde\alpha$ are generated by
\beq
\sum_{m=0}^r\tilde \alpha_{m}(r)t^m=\prod_{j=1}^{r}\big(1-(2j-1)^2t\big)\,,
\eeq
allows us to write (\emph{c.f.}, \eqn{I-L=2-Neven})
\bal
&\cI(2 r+1,2)= \frac{q^\frac{1}{3}q^{-\frac{(2 r+1)^2}{2}}}
{\eta^4(\tau)(2r)!(2r+1)! 2^{4r}} 
\sum_{m=0}^{r} \sum_{m' =0}^{r}\tilde\alpha_m(r)\tilde\alpha_{m'}(r)\\ 
& \hspace{1cm} \times \sum_{j=0}^\infty \sum_{n=0}^\infty 
(j+n+1)(2j+1)^{2r-2m} 
(2n+1)^{2r-2m'} q^{\frac{1}{2}(2 j+1)(2n+1)}\,.
\eal
In this case we are faced by double infinite sums
\bal
&\sum_{j=0}^\infty \sum_{n=0}^\infty(2j+1)^a(2n+1)^{a+2s+1} 
q^{\frac{1}{2}(2 j+1)(2n+1)}
\\&\qquad
=\frac{2^a\partial_\tau^a}{(\pi i)^a} \sum_{j=0}^\infty 
\sum_{n=0}^\infty(2n+1)^{2s-1} q^{\frac{1}{2}(2 j+1)(2n+1)}
= \frac{2^a\partial_\tau^a}{(\pi i)^a} H_{2s+1}\,, 
\eal
The quantities $H_{2s+1} = \sum_{n=0}^\infty(2n+1)^{2s+1} \frac{q^{n + \half}}{1-q^{2n+1}}$
also appeared in \cite{Zucker1979}. They are generated by
\beq
\sum_{s=0}^\infty(-1)^s H_{2s+1} \frac{t^{2s+1}}{(2s+1)!} = 
\frac{kK}{2 \pi} \sn\Big(\frac{2 K t}{\pi},k^2\Big)\,.
\eeq
Arbitrary numbers of $\tau$ derivatives of the $H_{2s+1}$ can again be 
straight forwardly evaluated using \eqn{tauderivatives}. 

This algorithm can be easily implemented to sum \eqn{I2largeN} for 
finite values of $N$. For $N=1,\cdots,4$ this gives
\begin{align}
\label{L=2finiteN}
\cI(1,2)&= \frac{q^{-\frac{1}{6}}k}{\eta^4(\tau)} 
\frac{K^2}{\pi^2} \,, 
\nn\\
\cI(2,2)&= 
\frac{q^{-\frac{5}{3}}}{\eta^4(\tau)} 
\frac{-3 E^2 K^2 + 2(2-k^2)E K^3 -(1-k^2)K^4}{6 \pi^4}\,,
\nn\\
\cI(3,2)&= \frac{q^{-\frac{25}{6}}k}{\eta^4(\tau)} 
\bigg(\frac{6E^2 K^4 -6(1-k^2)E K^5 +(1-k^2)^2 K^6}{12\pi^6} 
- \frac{E K^3 + k^2 K^4}{24 \pi^4} 
\nn\\&\hspace{2cm} 
+ \frac{K^2}{192 \pi^2}\bigg)\,,
\nn\\
\cI(4,2)&= \frac{q^{-\frac{23}{3}}}{\eta^4(\tau)} 
\bigg(\frac{-3 E^4 K^4 +4(2-k^2)E^3 K^5 -6(1-k^2)E^2 K^6 
+(1-k^2)^2 K^8}{72 \pi^8}
\nn\\ 
& \hspace{-1.2cm} + \frac{15 E^3 K^3 -15(2-k^2)E^2K^4 
+(11-11k^2 -4k^4)EK^5 +2(1-k^2)(2-k^2)K^6}{1080 \pi^6}
\nn\\ 
& \hspace{-1.2cm} - \frac{3E^2 K^2 -2(2-k^2)EK^3 +
(1-k^2)K^4}{432\pi^4} \Big).
\end{align}
The algorithm can easily be pushed to higher values of $N$ using Mathematica.

\subsection{Results from the $q$-expansion of longer quivers}
\label{sec:q}

In all the examples presented above the 
rescaled 
index 
$Z(N)$ \eqn{jacobiindexsun} 
at finite $N$ is expressed as a polynomial in $K$, $E$ and $k$. 
This is also true for the trivial case of arbitrary $L$ and $N=1$. 
This is just the theory of $L$ free hypermultiplets, where the index 
without flavour fugacities 
can be rewritten in terms of elliptic integrals as
\beq
\label{N=1index}
\cI(1, L>1)= 
\left(\frac{q^{-\frac{1}{12}}\vartheta_3} 
{\eta^{2}(\tau)\vartheta_2}\right)^LZ(1,1)=
\left(\frac{q^{-\frac{1}{12}}k^{\frac{1}{2}} K} 
{\pi\eta^{2}(\tau)}\right)^L.
\eeq
Inspired by these results,
we conjecture that for arbitrary $L$, $N$, 
the rescaled index $Z(N)$ 
is always given by a polynomial in complete elliptic 
integrals and the elliptic modulus%
\footnote{Note that 
$q^{\frac{L N^2}{4}} \frac{\vartheta_3^L \left( \frac{\pi \tau N}{2} \right)}
{\vartheta_3^L} = k^{\frac{L}{4}(1-(-1)^N)}$.}
\beq
\label{finiteNansatz}
Z(N,L)= 
k^{\frac{L}{2}(1-(-1)^N)} \sum_{j, l, m} a_{j,l,m}k^{2j} 
\left(\frac{K}{\pi}\right)^l\left(\frac{E}{\pi}\right)^m.
\eeq

Studying which terms appear in \eqn{finiteN-N=4}, \eqn{L=2finiteN} and
\eqn{N=1index} we guess that the only nonzero coefficients have
\bal
\label{constraints}
j\geq0\,, \quad l \geq L \,, \quad k&\geq0\,,\\ 
l-m-2j &\geq0\,, \\
LN-l-m &\geq0\hspace{0.6cm}\text{is even.} \\
\eal
These constraints leave us with finitely many $a_{j,l,m}$,
which we can fix by comparing the $q$ expansions of each side of \eqn{finiteNansatz}. 
We first use the relations \eqn{ISUN}, \eqn{grandpartition} and \eqn{SUNXi} to express 
the left hand side as
\beq
\label{qexpandable}
Z(N,L)= 
\sum_{\vec{n} \in \bZ^{L-1}} 
\prod_{p \in \bZ} \bigg(1 + \kappa\prod_{a=1}^L 
\frac{1}{q^{p-n^{(a)}+\half}+q^{-p+n^{(a)}-\half}}\bigg)
\bigg|_{\kappa^N} \,,
\eeq
where $|_{\kappa^N}$ indicates extracting the 
coefficient of $\kappa^N$. Now the $q$ expansion can be easily 
obtained by truncating the sum over $\vec{n}$ and the product 
over $p$ at large orders. 
Solving the resulting linear problems for the $a_{j,l,m}$ 
and reintroducing the scaling factor in \eqn{jacobiindexsun} 
we have obtained the results
\begin{align}
\label{guessed expressions}
\cI(2,3)&= \frac{q^{-\frac{5}{2}}}{\eta^6(\tau)}
\bigg(\frac{- E^3 K^3 +3 E^2 K^4 -3(1-k^2)E K^5 +(1-k^2)^2 K^6}{2 \pi^6}
-\frac{k^2 K^4}{8 \pi^4} \bigg),
\nn\\
\cI(3,3)&= \frac{q^{-\frac{25}{4}} k^{\frac{3}{2}}}{\eta^6(\tau)}
\bigg(\frac{-(1-k^2)^2(1+k^2)K^9}{120 \pi^9} - \frac{8 E K^4-(29+21 k^2)K^5}{1920 \pi^5}
\nn\\
& \hspace{2cm}- \frac{24 E^2 K^5 -24(1-k^2) E K^6 +5(1-k^2)^2 K^7}{96 \pi^7} + 
\frac{K^3}{1536 \pi^3}\bigg),
\nn\\
\cI(2,4)&=\frac{q^{-\frac{10}{3}}}{\eta^8(\tau)}
\bigg(\frac{-3 E^4 K^4 +4(2-k^2)E^3 K^5 -6(1-k^2)E^2 K^6 
+(1-k^2)^2 K^8}{6 \pi^8} 
\nn\\
& \hspace{2cm} -\frac{2(1-k^2 +k^4)EK^5 -(1-k^2)(2-k^2)K^6}{45 \pi^6} \bigg).
\end{align}
To fix a unique solution for the first, second and third equalities of
\eqn{guessed expressions} we required the $q$ expansions of 
\eqn{finiteNansatz} up to $q^{19}$, $q^{38}$ and $q^{38}$ respectively.
We have further checked that the solutions reproduce 
the $q$ expansions of the right hand side of \eqn{qexpandable} 
up to $q^{90}$, $q^{90}$ and $q^{48}$ respectively. 
One could continue to larger values of $N$ and $L$, but the number of 
terms required in \eqn{qexpandable} grows very quickly.

\section*{Acknowledgements}

We would like to thank
Mahesh Kakde, 
Dario Martelli, 
Sameer Murthy, 
Eric Verlinde 
and 
Kostya Zarembo
for discussions.
N.D. Would like to thank the hospitality of the IFT, Madrid, 
during the course of this work. 
The research of J.B. has received funding from the People Programme 
(Marie Curie Actions) of the European 
Union's Seventh Framework Programme FP7/2007-2013/ under REA 
Grant Agreement No 317089 (GATIS). 
The research of N.D. is underwritten by an STFC advanced fellowship.
The research of J.F. is funded by an STFC studentship ST/K502066/1.

\appendix

\section{The index of $\cN=2$ multiplets and theta functions}
\label{ellipticgamma}

The most general index of generic $\cN=1$ superconformal theory in 4d depends on three 
fugacities for space-time and $R$ symmetry, denoted by $p$, $q$ and $t$. 
The chiral multiplet with flavour fugacity $z$ is written as
\beq
\mathcal{I}^{\cN =1}_{\text{chir}} 
= \Gamma_e\left(tz;p^2,q^2\right), 
\eeq
where $\Gamma_e$ is the elliptic gamma function, defined by
\beq
\label{gammae}
\Gamma_e (z;r ,s)
= \prod_{j,k>0} 
\frac{1- z^{-1} r^{j+1} s^{k+1}} 
{1 - z r^j s^k}
\,.
\eeq
An $\cN =2 $ hypermultiplet then 
contributes the product of two elliptic gamma functions 
\beq
\label{n2hyp}
\mathcal{I}^{\cN=2}_{\text{hyp}} 
=\Gamma_e\left(tz;p^2,q^2\right)
\Gamma_e\left(tz^{-1};p^2,q^2\right).
\eeq

The Schur limit corresponds to $t=q$, and the equation above becomes
\beq
\label{schurhyp}
\mathcal{I}^{\cN=2}_{\text{hyp}} 
=\Gamma_e(qz;p^2, q^2)\Gamma_e(qz^{-1};p^2 , q^2)
=\frac{1}{\theta(q z,q^2)} \,.
\eeq
This last expression is a $q$-theta function defined as
\beq
\label{qtheta}
\theta(z,q) = \prod_{n=0}^\infty (1- z q^n ) (1- q^{n+1}/z)\,,
\eeq
and it is indeed simple to check from the definition \eqn{gammae} that the product of 
the two gamma functions in \eqn{schurhyp} reduce to a theta function.

As was already done in \cite{Dolan2009} for the $\cN=1$ case, the 
contribution from an $\cN = 2$ 
vector multiplet can also be expressed in terms of the $q$-theta function as
\beq
\label{N2vec}
\mathcal{I}^{\cN =2}_{\text{vec}} = 
\frac{(q^2;q^2)_\infty^{2(N-1)}}{N!}\frac{1}{\pi^N} 
\int_0^\pi
d^N \alpha 
\prod_{i<j}
\theta (e^{-2 i (\alpha_i - \alpha_j)} , q^2 )
\theta (e^{2 i (\alpha_i - \alpha_j)} , q^2 )
\,.
\eeq
The prefactor includes a $q$-Pochhammer symbol, defined for $|q|<1$ by
\beq
(a; q)_\infty = \prod_{r=0}^\infty(1- a q^r)\,.
\eeq
Clearly the $q$-theta function \eqn{qtheta} is the product of two $q$-Pochhammer symbols.

\section{Definitions and useful identities}
\label{identities}

In this paper we chose to use Jacobi theta functions and the Dedekind eta function rather 
than $q$-theta functions and $q$-Pochhammer symbols. These are related by
\beq
\label{qtojacobi}
\theta(e^{2 iz}, q^2)
= \frac{-i e^{iz}\,\vartheta_1(z, q)}{q^{1/6}\eta(\tau)} 
\,, 
\qquad \left(q^2; q^2 \right)_\infty 
= q^{-1/12} \eta(\tau)\,.
\eeq
where the (quasi)period $\tau$ is related to the nome $q$ by $q= e^{i\pi\tau} $.
The Jacobi theta function $\vartheta_3(z,q)$ is given by the series and product representations
\bal
\label{thetaseriesproduct}
\vartheta_3(z,q)&= 
\sum_{n=-\infty}^\infty q^{n^2} e^{2 i n z} = \prod_{k=1}^\infty 
\left(1- q^{2k} \right)\left(1 + 2 q^{2k-1} \cos \left(2 z \right)+ q^{4k-2} \right)\,, \\
\vartheta_2(z,q)&=q^{\frac{1}{4}} e^{-i z} \vartheta_3(z - \tfrac{1}{2} \pi\tau,q)
=\sum_{n=-\infty}^\infty q^{(n+\half)^2} e^{i(2n+1)z} 
\\& = 
2 q^{\frac{1}{4}} \cos(z)\prod_{k=1}^\infty \left(1- q^{2k} \right)
\left(1 + 2 q^{2k} \cos \left(2 z \right)+ q^{4k} \right).
\eal
The remaining two theta functions are given by
\bal
\label{auxtheta}
\vartheta_1(z,q)&= 
i q^{\frac{1}{4}} e^{- i z} 
\vartheta_3(z - \tfrac{1}{2} \pi\tau - \tfrac{1}{2} \pi,q)\,,
\\
\vartheta_4(z,q)&= \vartheta_3(z - \tfrac{1}{2} \pi,q)\,.
\eal

$\vartheta_3$ satisfies the quasi-periodic properties for any integers $n,m$
\beq
\label{quasiperiodicity}
\vartheta_3(z + n \pi + m \pi\tau, q)
= 
q^{- m^2} e^{-2i z m} \vartheta_3(z, q)\,.
\eeq
We also give here formulae to evaluate integrals of derivatives of theta functions
\bal
\label{integrals}
\frac{1}{2 \pi i} \int_{- i\pi}^{i\pi} d \mu e^{-m \mu} \partial_\mu^l \vartheta_3(i\mu, q)
&= m^l q^{\frac{m^2}{4}} \frac{1}{2}(1 +(-1)^m)\,, \\
\frac{1}{2 \pi i} 
\int_{- i\pi}^{i\pi} d \mu e^{-m \mu} \partial_\mu^l \vartheta_2(i\mu, q)
&= m^l q^{\frac{m^2}{4}}\frac{1}{2}(1 -(-1)^m)\,.
\eal
Jacobi's imaginary transformation with $\tau = -1/\tau'$, and $q' \equiv e^{i\pi\tau'}$ are
\bal
\label{Jacobiimaginary}
\vartheta_1(z, q)&=(-i)(-i\tau)^{-\half} e^{i\tau' z^2/\pi} \vartheta_1(\tau' z, q')\,,
\\
\vartheta_2(z, q)&=(-i\tau)^{-\half} e^{i\tau' z^2/\pi} \vartheta_4(\tau' z, q')\,,
\\
\vartheta_3 \left(z, q \right)&=(-i\tau)^{-\half}e^{i\tau' z^2/\pi} \vartheta_3(\tau' z, q')\,,
\\
\vartheta_4 \left(z, q \right)&=(-i\tau)^{-\half} e^{i\tau' z^2/\pi} \vartheta_2(\tau' z, q')\,.
\eal

We also use in the main text the formula
\beq
\label{reduction} 
\vartheta_3 \vartheta_2 \vartheta_4 = 2 \eta(\tau)^3 \,,
\eeq
as well as (see 20.7(iv) of \cite{Olver2010})
\beq
\label{reductionbis}
\eta^2(\tau/2)= \vartheta_4 \eta(\tau)\,.
\eeq
We also require Watson's identity (see 20.7(v) of \cite{Olver2010})
\beq
\label{Watson}
\vartheta_3(z, q)\vartheta_3(\omega, q)= 
\vartheta_3 \big(z + \omega, q^2 \big)\vartheta_3 \big(z - \omega, q^2 \big)
+ \vartheta_2 \big(z + \omega, q^2 \big)\vartheta_2 \big(z- \omega, q^2 \big)\,.
\eeq

\subsection*{An infinite sum in terms of Jacobi theta functions}
We make use in Section~\ref{largeNcomplexanalysis} of the formula
\beq
\label{cossinhseries}
\sum_{n=1}^\infty(-1)^{n} \frac{\cos 4 \alpha n}{n \sinh(- i\pi\tau n)} 
=-\frac{i\pi\tau}{12} - \frac{1}{6} \log \frac{4}{kk'} 
-\log \frac{\vartheta_3 \left(2 \alpha, q \right)}{\vartheta_3(0, q)}\,,
\eeq
which is a combination of (see 16.30.3 of \cite{Abramowitz1964})
\beq
\label{16.30.3} 
\sum_{n=1}^\infty \frac{(-1)^{n}}
{n \sinh(- i\pi\tau n)}(1 - \cos 4 \alpha n)
=\log \frac{\vartheta_3 \left(2 \alpha, q \right)}{\vartheta_3(0, q)} \,,
\eeq
and (see T1.3 of \cite{Zucker1984})
\beq
\label{T13}
\sum_{n=1}^\infty \frac{(-1)^n}
{n \sinh(- i\pi n \tau)} 
=
- \frac{i\pi\tau}{12} - \frac{1}{6} \log \frac{4}{kk'}\,,
\eeq
where the elliptic modulus and complementary elliptic modulus are respectively defined as
\beq
\label{elliptick}
k = \frac{\vartheta_2^2}{\vartheta_3^2} \,, 
\quad 
k' = \frac{\vartheta_4^2}{\vartheta_3^2} \,.
\eeq

\subsection*{A multiple angle formula for theta functions}

\label{productproof}
We prove here a formula for the product of theta functions shifted by roots of unity 
used in Section~\ref{polynomialasymptotics}%
\footnote{This formula can also be found (without proof) in \cite{Wolfram}.}
\bal
\label{productformula}
\prod_{j=1}^L \vartheta_3& \big(z+ \tfrac{L-2j +1}{2L} \pi, q\big)\\
&= 
\prod_{n=1}^\infty \prod_{j=1}^L(1-q^{2n})
(1 +2 q^{2n-1} \cos \big(2z + \tfrac{\pi}{L}(L+2j-1)\big)+ q^{4n-2})\\
& = \prod_{n=1}^\infty \prod_{j=1}^L(1-q^{2n})
\big(1 + e^{i\pi\frac{L-2j+1}{L} +2 i z} q^{2n-1} \big)
\big(1 + e^{-i\pi\frac{L-2j+1}{L} - 2 i z} q^{2n-1} \big)\\
&= \prod_{n=1}^\infty(1-q^{2n})^L 
\big(1 + e^{2 i L z}q^{L(2n-1)} \big)\big(1 + e^{-2 i L z}q^{L(2n-1)}\big)
\\
&= \vartheta_3(L z, q^L)\frac{\eta^L(\tau)}{\eta(L \tau)}\,.
\eal

\section{A determinant identity for Jacobi theta functions}
\label{sec:elipticcauchy}

A crucial identity for our analysis is the generalization of the Cauchy determinant 
identity to theta functions. For arbitrary $x_i,y_j,t$ with $i,j=1,\cdots, n$ we have the 
identity for $q$-theta functions \cite{Frobenius1879,Krattenthaler2005}
\beq
\label{newdet}
\det_{ij}\left(\frac{\theta(t x_j y_i)}{\theta(t)
\theta(x_j y_i)} \right)
= \frac{\theta(t x_1 x_2 \cdots x_n y_1 y_2 \cdots y_n)}
{\theta(t)} 
\frac{\prod_{i < j} x_j y_j \theta(x_i/x_j)\theta(y_i/y_j)}
{\prod_{i,j} \theta(x_j y_i)}\,,
\eeq
where we have used the notation $ \theta(z)= \theta(z,q^2)$.

One can recover the usual Cauchy identity by taking the limit $q \rightarrow0$, where 
$\theta(z)\rightarrow 1-z$. Taking also the limit $t \rightarrow \infty$ we find 
\beq
\det_{ij} \left(\frac{1}{1- x_i y_j}\right)= 
\frac{\prod_{i<j}(x_i - x_j)(y_j - y_i)}
{\prod_{i,j}(1 - x_i y_j)} \,,
\eeq
and the usual form of the Cauchy identity is recovered by taking 
$x_i\rightarrow \frac{1}{x_i}$.

In the study of indices we encounter a determinant closely related to \eqn{newdet}. 
Making the replacement $x_i\rightarrow e^{2 i\alpha_i}$, $y_i\rightarrow q e^{-2 i\alpha'_i}$ 
as well as $t \rightarrow -q^{2T}$, and rewriting the expression in terms of 
Jacobi theta functions yields
\begin{align}
&\frac{\prod_{i < j} 
\vartheta_1 \big(\alpha_i - \alpha_j \big)
\vartheta_1 \big(\alpha'_i - \alpha'_j \big)}
{\prod_{i,j=1}^N \vartheta_4 \big(\alpha_i - \alpha'_j \big)} 
\\&=
\det_{ij} \left(\frac{\vartheta_3 \big(\alpha_i - \alpha'_j + \pi\tau T \big)}
{\vartheta_4 \big(\alpha_i - \alpha'_j \big)} \right)
\frac{q^{-\frac{N^2}{4} - NT}} 
{\vartheta_2\big(\sum_{i=1}^N(\alpha_i - \alpha'_i)
+ \pi\tau(T + \frac{N}{2})\big)}
\frac{e^{- i N \sum_{i=1}^N(\alpha_i - \alpha'_i)}}
{\vartheta_2(\pi\tau T)^{N-1}} \,. 
\nn
\end{align}
By choosing $T= -\frac{1}{2}$ we obtain
\beq
\label{jacobidet}
\det_{ij} \left(\frac{\vartheta_2 \big(\alpha_i - \alpha'_j \big)}
{\vartheta_4 \big(\alpha_i - \alpha'_j \big)} \right)
\frac{q^{-\frac{N^2}{4}}}
{\vartheta_3\big(\sum_{i=1}^N(\alpha_i - \alpha'_i)+ N \tfrac{\pi\tau}{2} \big)}
\frac{e^{- i N \sum_{i=1}^N(\alpha_i - \alpha'_i)}}{\vartheta_3^{N-1}(0)} \,. 
\eeq

The ratio of Jacobi theta functions appearing in the determinant is in fact closely 
related to the Jacobi elliptic function $\cn$ 
\beq
\frac{\vartheta_2(z)} {\vartheta_4(z)}
= \frac{\vartheta_{2}}{\vartheta_{4}} \cn(z \vartheta_{3}^2)\,,
\eeq
where $\cn(z)\equiv \cn(z, k^2)$ and the elliptic modulus $k$ is defined in \eqref{elliptick}.

\section{An expansion formula}

Here we present a proof for
\beq
\label{expansionformula}
\hspace{-0.18cm}\left(\tfrac{\kappa + q + q^{-1} +
\sqrt{(\kappa + q + q^{-1})^2-4}}{2}\right)^{x} \hspace{-0.2cm} 
=\sum_{k=0}^\infty \sum_{l=0}^\infty 
\frac{x(x -k-1)!(x-l-1)!}{k!l!(x-k-l)!(x-k-l-1)!}
\kappa^{x-k-l}q^{k-l}\,.
\eeq
Our starting point is the expansion 
\beq
\label{expansion1}
\left(\frac{y +\sqrt{y^2-4}}{2}\right)^{x}= \sum_{s=0}^\infty 
\frac{(-1)^s x(x-s-1)!}{s!(x-2s)!} y^{x-2s}\,.
\eeq
Replacing $y^{x-2s} =(\kappa + q + q^{-1})^{x-2s}$
by its multinomial expansion gives
\beq
\sum_{s=0}^\infty\sum_{m=0}^\infty \sum_{j=0}^\infty 
\frac{(-1)^s x(x-s-1)!}{s!m!j!(x-2s-m-j)!}\kappa^{x-2s-m-j}q^{j-m}\,.
\eeq
Rewriting the sum in terms of indices $l=m+s$ and $k=j+s$ gives 
\beq
\sum_{s=0}^\infty\sum_{k=s}^\infty \sum_{l=s}^\infty 
\frac{(-1)^s x(x-s-1)!}{s!(k-s)!(l-s)!(x-k-l)!}\kappa^{x-k-l}q^{k-l}\,.
\eeq
Interchanging the order of summation we finally obtain
\bal
&\sum_{k=0}^\infty \sum_{l=0}^\infty \sum_{s=0}^{\text{min}(l,k)} 
\frac{(-1)^s x(x-s-1)!}{s!(k-s)!(l-s)!(x-k-l)!}\kappa^{x-k-l}q^{k-l} 
\\& \qquad
=\sum_{k=0}^\infty \sum_{l=0}^\infty 
\frac{x(x -k-1)!(x-l-1)!}{k!l!(x-k-l)!(x-k-l-1)!}\kappa^{x-k-l}q^{k-l}\,.
\eal

\addtolength{\parskip}{-.5mm}

\bibliographystyle{utphys2}

\bibliography{References}

\end{document}